\ifx\onecol\undefined
\documentclass[journal,twocolumn]{IEEEtran}
\else 
\documentclass[12pt,draftclsnofoot,journal,onecolumn]{IEEEtran}
\fi
\usepackage{amsfonts}
\usepackage{amsmath,amssymb}
\usepackage{acronym}  
\usepackage{algorithm}
\usepackage{algorithmic}
\usepackage{bm}
\usepackage{bbm}
\usepackage{booktabs}
\usepackage{color, soul}
\usepackage{cite}
\usepackage{graphicx}
\usepackage{indentfirst}
\usepackage{lipsum}
\usepackage{setspace}
\usepackage{tikz}
\usetikzlibrary{arrows}
\usepackage{subfigure}
\usepackage[amsmath,thmmarks]{ntheorem}
\usepackage{theorem}
\usepackage{url}

\newtheorem{proposition}{\bf Proposition}
\newtheorem{lemma}{\bf Lemma}

\newtheorem{remark}{\bf Remark}

\begin{document}

\title{Near-Field Wideband Beam Training Based on Distance-Dependent Beam Split}

\author{Tianyue~Zheng, Mingyao~Cui, Zidong~Wu, and Linglong~Dai, {\textit{Fellow, IEEE}}
\thanks{
This work was funded in part by the National Science Fund for Distinguished Young Scholars (Grant No. 62325106), and in part by the National Key R\&D Program of China (No. 2023YFB811503). 
}
\thanks{T. Zheng, Z. Wu, and L. Dai are with the Department of Electronic Engineering, Tsinghua University, Beijing 100084, China, and also with the Beijing National Research Center for Information Science and Technology (BNRist), Beijing 100084, China (e-mails: \{zhengty22, wuzd19\}@mails.tsinghua.edu.cn, \{daill\}@tsinghua.edu.cn).
M. Cui is with the Department of Electrical and Electronic Engineering, the University of Hong Kong, Hong Kong (e-mail: cuimy23@connect.hku.hk).
}}

\maketitle
\begin{abstract}
    Near-field beam training is essential for acquiring channel state information in 6G extremely large-scale multiple input multiple output (XL-MIMO) systems.
    To achieve low-overhead beam training, existing method has been proposed to leverage the near-field beam split effect, which deploys true-time-delay arrays to 
    simultaneously search multiple angles of the entire angular range in a distance ring with a single pilot.
    However, the method still requires exhaustive search in the distance domain, which limits its efficiency. 
    To address the problem, we propose a distance-dependent beam-split-based beam training method to further reduce the training overheads. 
    Specifically, we first reveal the new phenomenon of distance-dependent beam split, where by manipulating the configurations of time-delay and phase-shift, beams at different frequencies can simultaneously scan the angular domain in multiple distance rings. 
    Leveraging the phenomenon, we propose a near-field beam training method where both different angles and distances can simultaneously be searched in one time slot. Thus, a few pilots are capable of covering the whole angle-distance space for wideband XL-MIMO. Theoretical analysis and numerical simulations are also displayed to verify the superiority of the proposed method on beamforming gain and training overhead.
\end{abstract}


\vspace{-1em}
\section{Introduction} \label{sec-intro}
With the dramatically increasing demand for data transmission in future sixth-generation (6G) communications, a 10-fold increase in spectrum efficiency compared with 5G is anticipated in 6G~\cite{6G2021,6G2024}. To fulfill this vision, 5G massive multiple input multiple output (MIMO) systems will evolve into 6G extremely large-scale MIMO (XL-MIMO) systems~\cite{XLMIMO1,XLMIMO2,XLMIMO3}. Thanks to the dramatically increased antenna aperture, the spatial multiplexing and beamforming gains could be significantly enhanced in XL-MIMO~\cite{speeff}. 
To harvest these gains, acquiring accurate channel state information efficiently by beam training is necessary. Specifically, beam training is realized by selecting an optimal beam codeword from a predefined beam codebook so as to assign directional beams to target users, which serves as a dominate CSI acquition scheme in millimeter wave (mmWave) and terahertz (THz) XL-MIMO systems~\cite{BT1,BT2}.
    
\vspace{-1em}
\subsection{Prior Works}
\vspace{-0.2em}
Compared with 5G massive MIMO, beam training for 6G XL-MIMO will induce overwhelming training overheads due to fundamental change in electromagnetic (EM) field property. Specifically, in 5G systems when the antenna number at the base station (BS) is not very large, channel model mainly considers far-field propagation where radiated EM waves can be approximated as planar waves. Thus, the array response vector of the far-field channel is solely related to the angle. Correspondingly, the orthogonal Discrete Fourier Transform (DFT) codebook can conduct an efficient scan in the angular domain~\cite{exhaustive,DFT1,DFT2}, where each codeword in the codebook corresponds to a narrow beam directing to a specific angle.

As for the 6G XL-MIMO systems, however, the increased antenna aperture leads to more users located in spherical-wave-based near-field regions~\cite{nearCE,nearBF}. Unlike the angle-dependent far-field channel, the angle and distance from BS to user jointly
determine the array response vectors of near-field channels. Thereby, to perform near-field beam training, a polar-domain codebook was proposed in~\cite{nearCE} to simultaneously obtain accurate angle and distance information. 
This codebook partitions the entire “angle-distance” space into multiple grids, where each codeword is construced by the array response vector aligned with one sampled grid. 
During beam training, all codewords are sequentially tested, where the angle and distance information are estimated based on the codeword yielding the highest received power. Unfortunately, although this beam training method can acquire the near-field CSI accurately, it consumes unacceptable pilot overheads due to the two-dimensional exhaustive search for all possible BS-to-user angles and distances.

To cope with this problem, two categories of near-field beam training have been proposed, namely narrowband and wideband methods. Consider the narrowband methods at first. They are realized either by hierarchical beam training~\cite{hier_lu,hier_wei,hier_shi,AI_zhang} or inference from far-field beam training~\cite{far_you,far_qi,far_pan,farris_pan,hier_you}.
The near-field hierarchical beam training methods employ multi-resolution codebooks with different angle and distance coverages. 
Specifically, the spatial region covered by a lower-resolution codeword at any given layer is subdivided into several higher-resolution and non-overlapping spatial regions in the subsequent layer. During beam training, one can start from the lowest-resolution codebook and progress to
    the highest-resolution one layer by layer, and thereby gradually reduces the angle and distance estimation granularity by choosing the spatial region with larger received power in each layer~\cite{hier_lu,hier_wei,hier_shi,AI_zhang}. 
As for the inference from far-field beam training,  
    these methods attempt to recover near-field CSI with the aid of far-field codebook via a two-stage training strategy. In the first stage, the methods estimate all candidate angles based on the pattern of received power provided by far-field exhaustive beam training. Then the second stage uses a polar-domain codebook to search for distance within candidate angles or directly estimates the distance from far-field received beam pattern with deep learning-based methods~\cite{far_you,far_qi,far_pan,farris_pan,hier_you}. 

The above methods can to some extent reduce the training overheads.  Nevertheless, the narrowband methods only measures the received power of one subcarrier in a single time slot, limiting the information content obtained per pilot transition.  This constrains the efficiency of narrowband near-field beam training. In contrast, the majority of practical communication systems are wideband systems and it motivates us to develop wideband near-field beam training methods, which utilize information in multiple subcarriers~\cite{nearwide} to further reduce the beam training overheads. 
Unfortunately, to our best knowledge, few works have focused on wideband near-field beam training.

Excavating frequency resources, in our prior work~\cite{nearrainbow}, a wideband near-field beam training scheme, namely near-field rainbow, was designed based on near-field beam split. This scheme was built on the true-time-delay (TTD) array. By manipulating the time-delay (TD) parameters, multiple beams at different frequencies are
focused on multiple angles in a specific distance ring. In other words, multiple angles can be searched simultaneously in a single time slot, which significantly improves the training efficiency. However, since the beam split in~\cite{nearrainbow} is essentially distance-independent beam split, the optimal distance is obtained by sequentially traversing different distance rings. Thereby, the method in~\cite{nearrainbow} could not get rid of the additional overheads in the distance domain and results in relatively high training overheads.

\subsection{Our Contributions}
To tackle this problem, by exploiting \emph{distance-dependent} feature of near-field beam-split, we propose a near-field beam training method where both different angles and distances can be searched simultaneously in one time slot, achieving efficient beam training.\footnote{Simulation codes will be provided to reproduce the results in this paper: \url{http://oa.ee.tsinghua.edu.cn/dailinglong/publications/publications.html}.} 
Our contributions are summarized as follows.
\begin{itemize}
	\item Firstly, we reveal the mechanism of near-field distance-dependent beam split. Specifically,~\cite{nearrainbow} has proved that with the elaborate configuration of TD parameters, the beams at different frequencies could be made distributed on multiple angles of \emph{one-distance-ring}. In this paper, we prove that by introducing phase-shift (PS), the collaboration of TD and PS can spread the focused points of multi-frequency beams on different angles of  \emph{multi-distance-rings}. In this way, the traditional distance-independent beam split can be transformed into the distance-dependent beam split.
	\item Secondly, leveraging the mechanism of distance-dependent beam split, 
	an on-grid wideband beam training method is proposed to realize efficient near-field beam training. In the proposed scheme, both the angle and distance domain can be searched with beams of different subcarriers in a single pilot, where the beam with the highest power can be efficiently determined. Moreover, quantitative analysis of the required pilot number to guarantee the full coverage of the whole angle-distance space is presented.
	\item To further improve the beam training performance for off-grid users, we propose two off-grid beam training algorithms: the auxiliary beam pair-assisted method and match filter-based method. Their principle is to employ the envelope of received powers over multiple subcarriers, rather than relying on the single subcarrier with the highest power,  to jointly select the optimal codeword.
	\item 
    Finally, we present simulation results to validate the efficiency of the proposed method based on distance-dependent beam split. It shows that the proposed beam training scheme is capable of achieving near-optimal rate performance with a lower overhead compared to existing methods. Moreover, the enhanced algorithms tailored for off-grid users surpass the performance of the on-grid algorithm, particularly in regions of low signal-to-noise ratio (SNR).
\end{itemize}

\subsection{Organization and Notation}
\subsubsection{Organization}
The rest of the paper is organized as follows.
Section II introduces the near-field XL-MIMO wideband channel model, and formulate the near-field beam training problem. Then the mechanism of existing near-field distance-independent beam split will be reviewed. In Section III, the mechanism of distance-dependent beam split is introduced, and based on it we provide the design of an efficient wideband near-field beam training method. In Section IV, two improved beam training methods for off-grid users are described. Simulation results and conclusions are provided in Section V and Section VI, respectively.

\subsubsection{Notation}
${\bf a}^H$, ${\bf A}^H$ denote the conjugate transpose of vector $\bf a$ and matrix $\bf A$, respectively; $\left\| {\bf a} \right\|$ denotes the $l_2$ norm of vector $\bf a$; $\odot$ denotes Hadamard product; $\mathcal{C} \mathcal{N} ({\bf \mu},{\bf \Sigma})$ denotes the probability density function of complex multivariate Gaussian distribution with mean ${\bf \mu}$ and variance ${\bf \Sigma}$; ${\rm round}(a)$, $\lfloor a \rfloor$ and $\lceil a \rceil
$ denote rounding to, up to, and down to the nearest integer, respectively.

\section{System Model} \label{Sec_2}
In this section, we first introduce the near-field wideband
channel model in XL-MIMO systems and formulate the near-field beam training problem. Then, we will briefly review the traditional near-field beam split proposed in~\cite{nearrainbow}.

\subsection{Near-Field Wideband Channel Model} \label{Sec_2_Subsec_1} 
We examine a millimeter-wave wideband near-field XL-MIMO system. The BS equipped with an $N_t$-antenna uniform linear array serves a single-antenna user using orthogonal frequency division multiplexing (OFDM) of $M$ subcarriers. 
The bandwidth, central carrier frequency, speed of light, central wavelength, and antenna spacing are denoted as $B$, $f_c$, $c$, $\lambda_c=c/f_c$ and $d=\lambda_c/2$, respectively. The lower and upper subcarrier frequencies can be written as $f_L = f_c- B/2$ and $f_H = f_c+ B/2$ and the frequency of the $m$-th subcarrier as $f_m = f_c + \frac{B}{M}(m-1-\frac{M-1}{2})$. 
Besides, the number of BS antennas is assumed to be odd and is denoted by $N_t = 2N + 1$ for expression simplicity. 

In this paper, we mainly focus on the downlink beam training process. Let $s_m = 1$ be the normalized transmitted symbol at $f_m$, then the received signal $y_m$ at the user equipment (UE) of the $m$-th subcarrier is given by
\begin{equation} \label{eq-signal}
	y_m = \sqrt{P_t} {\bf h}_m^H  {\bm w}_m s_m+ n_m,
\end{equation}
where $P_t>0$ is the transmit power, ${\bf h}_m \in \mathbb{C}^{N_t} \times 1$ the downlink channel, ${\bm w}_m \in \mathbb{C}^{N_t \times 1}$ the unit-norm beamformer,  and  $n_m \sim \mathcal{CN}(0,\sigma^2)$ the complex circularly-symmetric Gaussian noise  with a variance of $\sigma^2$. We adopt the Saleh-Valenzuela channel model in~\cite{channel} to represent the downlink channel ${\bf h}_m \in \mathbb{C}^{N_t \times 1}$ at the $m$-th subcarrier as
\begin{equation}\label{eq_channel}
	{\bf h}_m = \sqrt{\frac{N_t}{L}} \sum_{l=1}^{L} \beta_{m,l} e^{-j2 \pi f_m \tau_l} {\bf a}_m(\theta_{l}, r_{l}),
\end{equation}
where $L$ is the number of paths being either line-of-sight (LoS)  or non-line-of-sight (NLoS) path, and $\beta_{m,l}$, $\tau_l$, $\theta_{l}=\sin \vartheta_l $, $r_{l}$ denote the path gain, time delay, spatial direction, and distance from the UE/scatter to the center of the antenna array of the $l$-th path, respectively.
Due to the significant path loss experienced by scattered signals, millimeter-wave communications heavily rely on the LoS path~\cite{8901159}, which is our primary focus:
\begin{equation}\label{eq_los}
	{\bf h}_m = \sqrt{N_t} \beta_{m} e^{-j k_m r} {\bf a}_m(\theta, r), 
\end{equation}
where $k_m = 2 \pi f_m/ c$ denotes the wavenumber at $m$-th subcarrier and the path gain of the LoS path is modeled as~\cite{8732419} 
\begin{equation}
	\beta_{m} = \frac{\lambda_m}{4 \pi r}.
\end{equation}
Therefore, we can obtain the relationship between the path gain at $m$-th subcarrier and central frequency $f_c$ as
\begin{equation}
	\beta_{m} = \frac{f_c}{f_m} \beta_{c}. 
\end{equation}
The near-field array response vector ${\bf a}_m(\theta, r)$ in (\ref{eq_los}) is derived based on spherical wave propagation:
\begin{equation}
	{\bf a}_m(\theta, r) = \frac{1}{\sqrt{N_t}} [e^{-j k_m (r^{(-N)}-r)},\ldots , e^{-j k_m (r^{(N)}-r)}]^T
\end{equation}
where $r^{(n)}=\sqrt{r^2+n^2 d^2-2 r \theta n d}$ represents the distance from the $n$-th BS antenna to the UE. By utilizing the Taylor expansion $\sqrt{1+x} \approx 1+\frac{1}{2}x-\frac{1}{8}x^2$~\cite{nearCE}, the distance difference $r^{(n)}-r$ can be approximated by:
\begin{equation}
	r^{(n)}-r \approx - nd\theta +n^2 d^2 \frac{1-\theta^2}{2r}.              
\end{equation}
For expression simplicity, we denote $\alpha = \frac{1-\theta^2}{2r}$. For fixed $\alpha$, the curve plotted by $(r = \frac{1-\theta^2}{2\alpha}, \theta)$ can be viewed as a distance ring in the physical space. In this case, ${\bf a}_m(\theta, r)$ can be approximated with ${\bf b}_m(\theta, \alpha)$ written as 
\ifx\onecol\undefined
\begin{equation}
	\begin{split}
		&{\bf a}_m(\theta, r) \approx {\bf b}_m(\theta, \alpha) = \\  & \frac{1}{\sqrt{N_t}} [e^{jk_m((-N)d\theta-(-N)^2 d^2 \alpha )},\ldots , e^{jk_m(Nd\theta-N^2 d^2 \alpha )} ].
	\end{split}
\end{equation}
\else 
\begin{equation}
		{\bf a}_m(\theta, r) \approx {\bf b}_m(\theta, \alpha) = \frac{1}{\sqrt{N_t}} [e^{jk_m((-N)d\theta-(-N)^2 d^2 \alpha )},\ldots , e^{jk_m(Nd\theta-N^2 d^2 \alpha )} ].
\end{equation}
\fi
Vector ${\bf b}_m(\theta, \alpha)$ effectively captures the spherical wave propagation in wideband scenarios.

\subsection{Formulation of Near-Field Beam Training Problem}
The near-field beam training attempts to estimate the location information $(\theta, \alpha)$ of the user, thereby aligning the beamforming vector with this location. To adapt to the near-field communication scenario, the optimal beamforming vector is typically selected from the predefined polar-domain codebook~\cite{nearCE}, which can be  represented as
\begin{equation}
	{\mathcal{A}}_m=[{\bf b}_m(\theta_1, \alpha_1^{1}),\cdots,{\bf b}_m(\theta_1, \alpha_1^{S_1}),\cdots,{\bf b}_m(\theta_{N_t}, \alpha_{N_t}^{S_{N_t}})],
\end{equation}
where $S_n$ is the number of sampled distance grids at $\theta_{n}$.
Each column of ${\mathcal{A}}_m$ is a codeword focused on the grid $(\theta_{n},\alpha_n^{s_n})$, with $s_n = 1,2, \cdots, S_n$. When conducting exhaustive beam training, we sequentially employ the codewords to perform beamforming and select the codeword leading to the largest received power to estimate the physical location $(\hat{\theta},\hat{\alpha})$, i.e.
\begin{equation}
	(\hat{\theta},\hat{\alpha}) = {\rm arg} \max_{(\theta_n,\alpha_n^{s_n})} \sum_{m=1}^M \| \sqrt{P_t} {\bf h}_m^H  {\bf b}_m(\theta_n, \alpha_n^{s_n}) + n_m\|^2.
\end{equation}
Finally, we set the beamforming vector as ${\bm w}_m(\hat{\theta},\hat{\alpha})={\bf b}_m(\theta_n, \alpha_n^{s_n})$ to serve the user. 
Consequently, the size of the polar-domain codebook is determined by the multiplication of the number of sampled angular grids and sampled distance grids. This results in a large codebook size, which in turn leads to prohibitive beam training overheads.

\subsection{Near-Field Distance-Independent Beam Split} \label{Sec_2_Subsec_2}
In this subsection, we review the mechanism of near-field beam split proposed in~\cite{nearrainbow} and explain how to employ it to achieve efficient beam training.

\ifx\onecol\undefined
\begin{figure}
	\centering 
	\includegraphics[width= \linewidth]{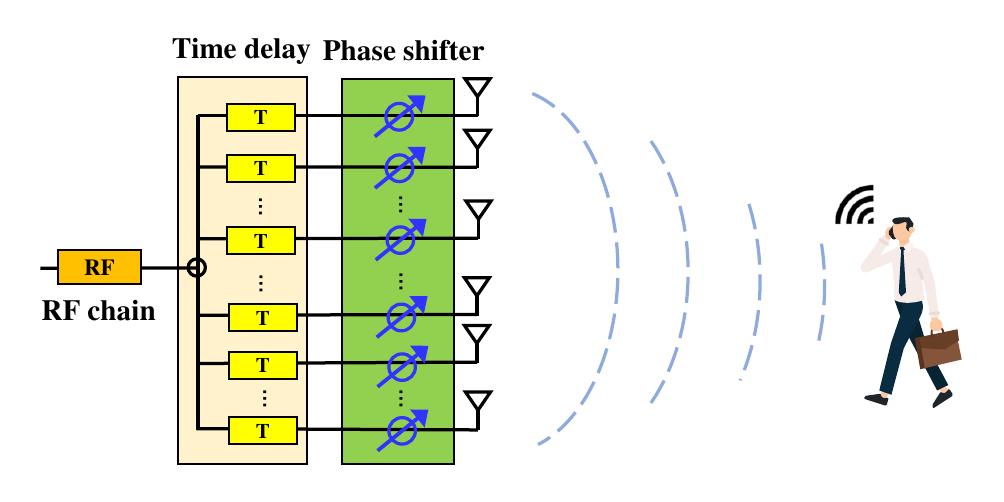}
        \caption{TD-PS precoding structure.}
	\label{pic_precoding}
\end{figure}
\else 
\begin{figure}
	\centering 
	\subfigure[Fully-digital TD-PS precoding structure.]{
		\includegraphics[width=0.5 \linewidth]{figures/TD_PS.pdf}} 
	\subfigure[Fully-digital TD precoding structure.]{
		\includegraphics[width=0.4 \linewidth]{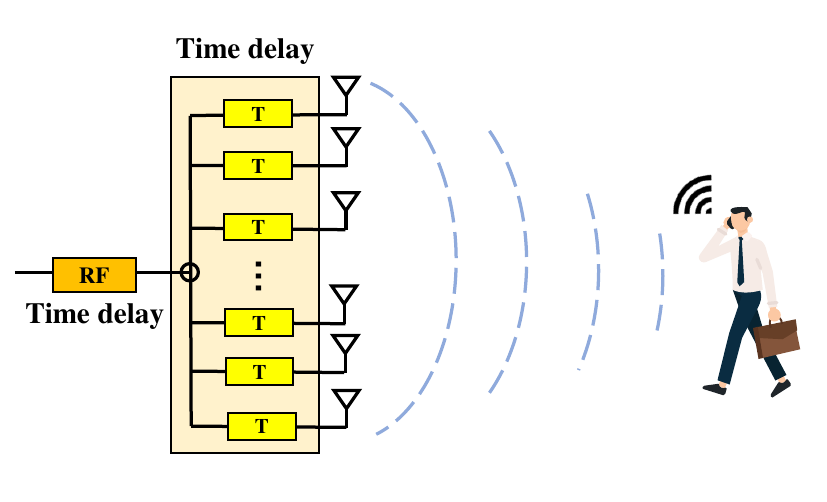}}
	\caption{Comparison of TD-PS precoding structure and TD precoding structure.}
	\label{pic_precoding}
\end{figure}
\fi

We utilize a TD-PS precoding structure, as presented in Fig.~\ref{pic_precoding}, to manipulate the near-field beam split effect.
Note that the TD array adopted in~\cite{nearrainbow} is a special case of TD-PS array by forcing the phase shift to zero. Therefore, the TD-PS array can achieve much more flexible control of the beam split effect, particularly in the distance dimension. 

As illustrated in Fig.~\ref{pic_precoding}, each antenna is sequentially connected to a time-delay unit and a phase-shift unit. A TD unit is capable of tuning {\it frequency-dependent} phase shift by actively delaying the signal. Accordingly, we can express the $n$-th element of the TD beamforming vector ${\bf w}_m^{\rm TD}$ at $f_m$, as
\begin{equation}\label{tdd1}
	[{\bf w}_m^{\rm TD}]_n = \frac{1}{\sqrt{N_t}} e^{-j 2 \pi f_m \tau^{(n)'} } = \frac{1}{\sqrt{N_t}} e^{-j k_m r_t^{(n)'} }
\end{equation}
where $\tau^{(n)'}$ is the adjustable delay of the $n$-th TD unit, while $r_t^{(n)'}\overset{\Delta}{=}c\tau^{(n)'}$. We use the superscript $'$ to indicate the adjustable parameters. Since ${\bf w}_m^{\rm TD}$ has a similar form to the array response vector ${\bf b}_m(\theta, \alpha)$, we can set $r_t^{(n)'}$ as $r_t^{(n)'}=nd\theta_t' -n^2 d^2 \alpha_t'$, where $\theta_t'$ and $\alpha_t'$ are defined as the \emph{adjustable TD parameters}. Thereby, the beamforming vector at $f_m$ can be rewritten as
\begin{equation}\label{tdd2}
	[{\bf w}_m^{\rm TD}(\theta_t',\alpha_t')]_n = \frac{1}{\sqrt{N_t}} e^{-j k_m (nd\theta_t'-n^2d^2\alpha_t') }.
\end{equation}
Furthermore, the $n$-th element of the beamforming vector, ${\bf w}^{\rm PS}(\theta_p',\alpha_p')$, generated by {\it frequency-independent} PSs can be similarly set as 
\begin{equation}
	[{\bf w}^{\rm PS}(\theta_p',\alpha_p')]_n = \frac{1}{\sqrt{N_t}} e^{-j k_c (nd\theta_p'-n^2d^2\alpha_p') },
\end{equation}
where $\theta_p',\alpha_p'$ are correspondingly \emph{adjustable PS parameters}.
By configuring TD-PS beamforming structure with parameters $\theta_t',\alpha_t',\theta_p',\alpha_p'$, the array gain at the $m$-th subcarrier on an arbitrary location $(\theta, \alpha)$ can be presented as
\ifx\onecol\undefined
\begin{eqnarray}\label{eq_gain}
	&&|({\bf w}_m^{\rm TD}(\theta_t',\alpha_t') \odot  {\bf w}^{\rm PS}(\theta_p',\alpha_p'))^T {\bf b}_m(\theta, \alpha)| \nonumber \\  &= & \frac{1}{N_t} \left\lvert \sum_{n=-N}^{N} e^{jnd(k_m (\theta- \theta_t') -k_c \theta_p' )-jn^2d^2(k_m (\alpha-\alpha_t')-k_c \alpha_p')} \right\rvert \nonumber \\ &=& G(k_m \theta-k_m \theta_t'-k_c \theta_p',k_m \alpha-k_m \alpha_t'-k_c \alpha_p'),
\end{eqnarray}
\else 
\begin{eqnarray}\label{eq_gain}
		|({\bf w}_m^{\rm TD}(\theta_t',\alpha_t') \odot  {\bf w}^{\rm PS}(\theta_p',\alpha_p'))^T {\bf b}_m(\theta, \alpha)|  &=& \frac{1}{N_t} \left\lvert \sum_{n=-N}^{N} e^{jnd(k_m \theta-k_m \theta_t'-k_c \theta_p' )-jn^2d^2(k_m \alpha-k_m \alpha_t'-k_c \alpha_p')} \right\rvert \nonumber \\ &=& G(k_m \theta-k_m \theta_t'-k_c \theta_p',k_m \alpha-k_m \alpha_t'-k_c \alpha_p'),
\end{eqnarray}
\fi
where we define $G(x,y)= \frac{1}{N_t} \left\lvert \sum_{n=-N}^{N} e^{jndx-jn^2d^2y} \right\rvert$ and it achieves a maximal value at $(x,y)=(0,0)$. According  to~\cite{nearrainbow}, the beam at frequency $f_m$ is focused on the location $(\theta_m,\alpha_m)$ with the maximum array gain. Thereby we obtain $k_m \theta_m-k_m \theta_t'-k_c \theta_p'=0$ and $k_m \alpha_m-k_m \alpha_t'-k_c \alpha_p'=0$, or equivalently:
\begin{align}
    \theta_m &= \theta_t' + \frac{f_c}{f_m} \theta_p', \label{eq_l1}\\
    \alpha_m &= \alpha_t' + \frac{f_c}{f_m} \alpha_p'. \label{eq_l2}
\end{align}
By setting $\theta_p'\neq 0$, the spatial angle $\theta_m$ is related to the frequency $f_m$, making beams at different frequencies split towards different spatial angles. Moreover, when $\alpha_p'=0$, $\alpha_m$ becomes independent of $f_m$, indicating that all beams are focused on the same distance ring $\alpha_m = \alpha_t'$. To perform fast beam training, the authors in \cite{nearrainbow} propose to distribute multi-frequency beams over multiple angles in the same distance ring simultaneously by setting $\theta_p' \neq 0$ and $\alpha_p'=0$. Then, it exhaustively searches different distance rings by configuring the value of $\alpha_t'$. 

However, it is clear that the near-field beam split in~\cite{nearrainbow} is essentially distance-independent, where beams can cover multiple angles in only {\it one} specific distance ring. It fails to fully exploit the degree-of-freedoms provided by the beam split effect, which still requires exhaustive search in the distance domain and limits its efficiency. To deal with this problem, we will introduce the distance-dependent beam split phenomena and how we can search multiple angles as well as distances at the same time with a single pilot by utilizing this phenomenon.

{

\section{Distance-Dependent Beam Split and Proposed Near-Field Beam Training Scheme } \label{proposed}
In this section, the mechanism of distance-dependent beam split is elaborated. Thanks to the periodicity of the angle dimension, beams can skip periodically between $\theta=1$ and $\theta=-1$. Therefore, by elaborately designing the TD and PS parameters, we can divide the subcarriers over the entire bandwidth into several groups, and the beams of each group cover the angular range with different distance ranges.
Taking advantage of this mechanism, we propose a distance-dependent beam-split-aided near-field beam training method. In the proposed method, 
multiple distance rings can be covered by the beams at different frequency groups simultaneously in the near field. Based on the discovery, both angle and distance dimensions can be searched simultaneously with different subcarriers in a single pilot during beam training.

\subsection{Distance-Dependent Beam Split}\label{phenomena}
For a better understanding of the mechanism of distance-dependent beam split, in this subsection, we start from the periodical beam split in the far field. Then we will illustrate the distance-dependent beam split in near-field scenarios. 

\subsubsection{Mechanism of Far-field Periodical Beam Split}\label{sec:Far-field-beam-split}
To illustrate the far-field periodical beam split phenomena, we first analyze the periodicity of the array gain with the TD-PS precoding structure. Specifically, the distance $r$ is assumed to be larger than the Rayleigh distance so that distance-related parameters $\alpha$ reliably approach 0. Then, the array response vector becomes $[{\bf b}_m(\theta, \alpha)]_n=[{\bf b}_m(\theta,0)]_n=\frac{1}{\sqrt{N_t}} e^{jk_mnd\theta}$. Accordingly, the TD and PS beamforming vector can be set as $[{\bf w}_m^{\rm TD}(\theta_t',0)]_n= \frac{1}{\sqrt{N_t}} e^{-j k_m nd\theta_t'}$ and $[{\bf w}^{\rm PS}(\theta_p',0)]_n = \frac{1}{\sqrt{N_t}} e^{-j k_c nd\theta_p' }$, respectively. Therefore, the array gain in (\ref{eq_gain}) can be simplified into
\begin{equation}
    G(k_m \theta-k_m \theta_t'-k_c \theta_p',0) = \Xi_{N_t}\left(\frac{d k_m}{\pi}(\theta - \theta_t' - \frac{f_c}{f_m}\theta_p')\right),
\end{equation}
where $\Xi_{N_t}(x)= \sin \frac{N_t \pi}{2}x/N_t \sin \frac{\pi}{2} x$ is the Dirchlet sinc function.
The far-field array gain $G(x, 0)$ is evidently a periodic function of period $2\pi/d$, i.e., 
\ifx\onecol\undefined
\begin{align} \label{eq_peri_far}
	G(x-\frac{2p\pi}{d}, 0) =  G(x, 0), \:\:\forall p \in \mathbb{Z}.
\end{align}
\else 
\begin{equation} \label{eq_peri_far}
	G(x-\frac{2p\pi}{d}, 0) = \left\lvert \sum_{n=-N}^{N} e^{jnd(x-\frac{2p\pi}{d})} \right\rvert =\left\lvert \sum_{n=-N}^{N} e^{jndx-j2\pi np} \right\rvert= G(x, 0).
\end{equation}
\fi
This periodicity implies that all the points $x=\frac{2p\pi}{d}, p \in \mathbb{Z}$ can maximize the array gain $G(x, 0)$. In this context, the spatial angle $\theta_m$ at frequency $f_m$ should be derived from
$
	k_m \theta_m -k_m \theta_t'-k_c \theta_p' = \frac{2p_m \pi}{d},
$
as 
\begin{equation}\label{eq_far}
	\theta_m = \theta_t' + \frac{f_c}{f_m} (\theta_p'+2p_m).
\end{equation}
The physical significance of $p_m$ in \eqref{eq_far} is interpreted as follows.
\begin{remark} (Physical significance of integer $p$): 
	Since the spatial angle $\theta_m$ corresponds to an actual physical angle $\vartheta$, $\theta_m$ has to satisfy the angle constraint $\theta_m \in [-1,1]$. To realize the case of $p_m=0$ as presented in \eqref{eq_l1}, the variables $\theta_t'$ and $\theta_p'$ need to be carefully designed to make $\theta_m = \theta_t' + \frac{f_c}{f_m} \theta_p'$ satisfy the angle constraint $\theta_m \in [-1,1]$. 
    However, as long as the delay circuits allow, the variables $\theta_t'$ and $\theta_p'$ can be arbitrarily adjusted, making it possible to break the angle constraint: $\theta_t' + \frac{f_c}{f_m} \theta_p' \notin [-1,1]$. In this context, the integer $p_m$ is activated and it 
    will naturally become non-zero to shift the value of $\theta_m = \theta_t' + \frac{f_c}{f_m} (\theta_p'+2p_m)$ into the physical  angle range, $[-1, 1]$. In other words, by adjusting the delay parameters $\theta_t'$ and $\theta_p'$, the integer $p_m$ can be automatically controlled to satisfy the angle constraint.
\end{remark}

\eqref{eq_far} allows us to illustrate the far-field periodical beam split. 
To begin with, we adjust the parameters $\theta_t'$ and $\theta_p'$ such that the beam of the first subcarrier is steered to $\theta_1 = \theta_t' + \frac{f_c}{f_1}\theta_p'=-1$ with $p_1=0$. 
Without loss of generality, we assume $\theta_t' > 0$ and $\theta_p' < 0$. In this case, as the subcarrier frequency $f_m$ grows, the focused direction $\theta_m$ gradually increases from $-1$ to $1$ with $p_m$ remaining to $0$. Suppose that the value of $\theta_p'$ is large enough such that the focused direction $\theta_m$ can approach $\theta_{m_1} =\theta_t' + \frac{f_c}{f_{m_1}}\theta_p'= 1$ at the $m_1$-th subcarrier
where $m_1 < M$. Subsequently, if we continue to increase the frequency, then $\theta_{m_1+1} = \theta_t' + \frac{f_c}{f_{m_1 + 1}}\theta_p'$ definitely exceeds $1$. According to \textbf{Remark 1}, the integer $p_{m_1+1}$ will automatically become $-1$ to shift $\theta_{m_1+1}$ from $\theta_t' + \frac{f_c}{f_{m_1 + 1}}\theta_p'$ to $\theta_t' + \frac{f_c}{f_{m_1 + 1}}(\theta_p' - 2)$ for guaranteeing the angle constraint $\theta_{m_1 + 1} \in [-1,1]$. Besides,  the beam at frequency $f_{m_1+1}$ is steered to $\theta_t' + \frac{f_c}{f_{m_1+1}} (\theta_p'-2) \approx -1$. That is to say, due to the periodic property of the angle domain, the spatial direction of the beam ``skips" from $1$ to $-1$
when the frequency arises from $f_{m_1}$ to $f_{m_1 + 1}$. 
In addition, with the continuous increment of the frequency, the focused direction of the beam will gradually rise until the focused direction $\theta_{m_2}$, with $m_1<m_2<M$,  arrives $1$ for the second time. Then, the beam will “skip” again to $-1$ by automatically tuning the integer $p_{m_2 + 1}$ to $-2$. This beam skip process will be repeated until the last subcarrier $f_M$\footnote{If TD-PS parameters are initialized as $\theta_p'+2p>0$, the focus direction of the beam will decrease conversely with respect to frequency $f_m$. Then the beam will skip from $-1$ to $1$ when automatically increasing $p$.}. This phenomenon is called the {\it periodical beam split}.

\begin{remark}
	Utilizing far-field periodical beam split, an important insight is that several beams of different frequencies $f$ may focus in the same physical direction. In other words, the beams over the entire bandwidth can periodically scan the angle range $[-1,1]$.
\end{remark}
}

{
\subsubsection{Mechanism of Near-field Distance-Dependent Beam Split}
Consider the periodic pattern of beam split in the polar domain. 
Similar to the periodic far-field array gain in \eqref{eq_peri_far}, the near-field array gain $G(x,y)$ exhibits a periodicity as well. As proven in \cite{nearrainbow}, $G(x,y)$ is periodic 
against the vector variable $(x,y)$ with a period of $(\frac{2\pi}{d},\frac{2\pi}{d^2})$, i.e., $G(x-\frac{2\pi p}{d},y-\frac{2\pi q}{d^2}) = G(x,y)$, for $\forall p,q \in \mathbb{Z}$.
The periodicity implies that $(x,y)=(\frac{2\pi p}{d},\frac{2\pi q}{d^2})$ are all optimal solutions to maximize the array gain $G(x, y)$. Then the focused location $(\theta_m,\alpha_m)$ of the beam at frequency $f_m$ in \eqref{eq_l1} and \eqref{eq_l2} should be rewritten as 
\begin{align}
    \theta_m &= \theta_t' + \frac{f_c}{f_m}(\theta_p'+ 2p_m), \label{eq_change} \\
    \alpha_m &= \alpha_t' + \frac{f_c}{f_m}(\alpha_p'+\frac{2q_m}{d}), \label{eq_alpha}
\end{align}
which are derived from $k_m \theta-k_m \theta_t'-k_c \theta_p'=\frac{2\pi p_m}{d}$ and $k_m \alpha-k_m \alpha_t'-k_c \alpha_p'=\frac{2\pi q_m}{d^2}$. 
Clearly, the periodic beam split pattern in the near field is a generalization of \eqref{eq_far} in the far field.
When $B \ll f_c$, \eqref{eq_change} and \eqref{eq_alpha} can be approximated by the first-order Taylor expansion as  
\begin{align}
    \theta_m &\approx  \theta_t'+2 (\theta_p'+2p)  - \frac{ \theta_p'+2p }{f_c} f_m, \label{eq_appro} \\
    \alpha_m &\approx  \alpha_t'+2 (\alpha_p'+\frac{2q_m}{d})  - \frac{\alpha_p'+\frac{2q_m}{d} }{f_c} f_m.\label{eq_alpha_appro}
\end{align}
\eqref{eq_appro} and \eqref{eq_alpha_appro} suggest that the focused direction/distance of the beam grows linearly with the frequency $f_m$. 
}

{
We summarize the mechanism of near-field distance-dependent beam split as follows.  
Formulas \eqref{eq_change} and \eqref{eq_alpha} reveal that TD-PS  arrays can control the coverage range of multi-frequency beams over the \emph{angular} and \emph{distance} dimensions simultaneously. 
Consider the angular dimension at first. The formula \eqref{eq_change} has the same expression as \eqref{eq_far} in the far-field case. The beam split pattern over the angular dimension in near-field scenarios is identical to that in the far field. Thereafter, the manipulation of parameters $\theta_t'$ and $\theta_p'$ can be the same as what we have discussed in Section~\ref{sec:Far-field-beam-split}.
We then consider the focused distance of multi-frequency beams presented in \eqref{eq_alpha}.
The major difference between the beam split pattern over the angular and distance dimensions is attributed to the physical constraints imposed on $\theta_m$ and $\alpha_m$. The physical direction $\theta_m$ is bounded between $-1$ and $1$ while the distance ring, $\alpha_m$, can grow from 0 to infinity. This difference implies that, as opposed to the periodic pattern of the angle-dimension beam split, the focused distance ring $\alpha_m$ increases/decreases monotonically with the frequency $f_m$, as long as $\alpha_p'+\frac{2q_m}{d} \neq 0$ and $\alpha_t' + \frac{f_c}{f_m}(\alpha_p'+\frac{2q_m}{d}) \ge 0$. 
Moreover, we can just ignore the subscript $m$ of $q_m$ because $q_m$ remains the same when changing the frequency $f_m$. 
For example, if the multi-frequency beams are desired to cover the entire distance range $[\alpha_{p},\alpha_{q}]$, we can simply set
\begin{align} 
    \alpha_L &= \alpha_t' + \frac{f_c}{f_L} (\alpha_p'+\frac{2q}{d})
	\geqslant  \alpha_{q},\label{eq_alphaL} \\
    \alpha_H &= \alpha_t' + \frac{f_c}{f_H} (\alpha_p'+\frac{2q}{d})  \leqslant \alpha_{p}, \label{eq_alphaH}
\end{align}
where $\alpha_p'+\frac{2q}{d}$ is assumed to be greater than 0 without loss of generality. 

In summary, by integrating the periodic pattern of the angle-dimension and the monotonous pattern of the distance-dimension, we arrive at the distance-dependent beam split effect. As presented in Fig.~\ref{pic_skip}(a), the focused directions of the multi-frequency beams fluctuate periodically between $[-1,1]$ while the focused distance ring increases monotonically with the frequency $f_m$. In such cases, the focused locations $(\theta_m,\alpha_m), m=1,2 \ldots M$ are distributed over several inclined strips. Each strip covers the entire angle range $[-1,1]$ while different strips occupy different distance ranges.
}


\ifx\onecol\undefined
\begin{figure*}
	\centering 
	\subfigure[]{
		\includegraphics[width=0.3 \linewidth]{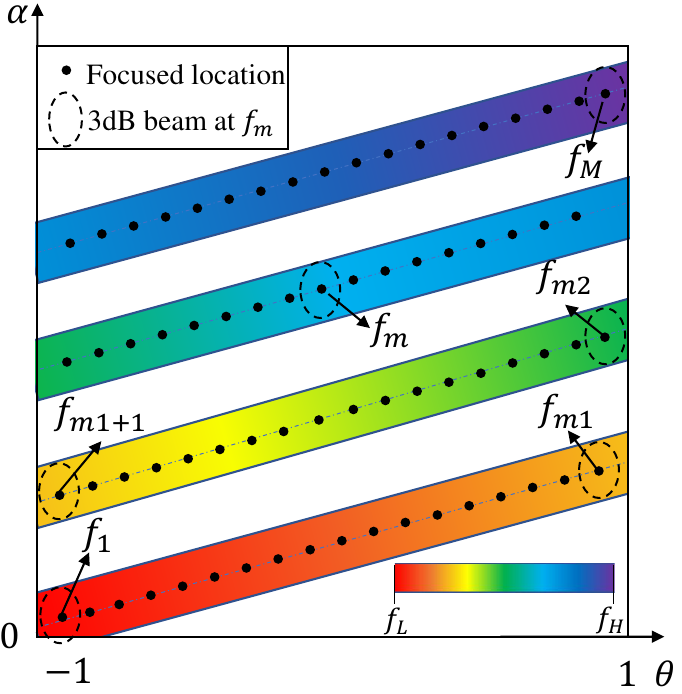}} 
        \subfigure[]{
		\includegraphics[width=0.31 \linewidth]{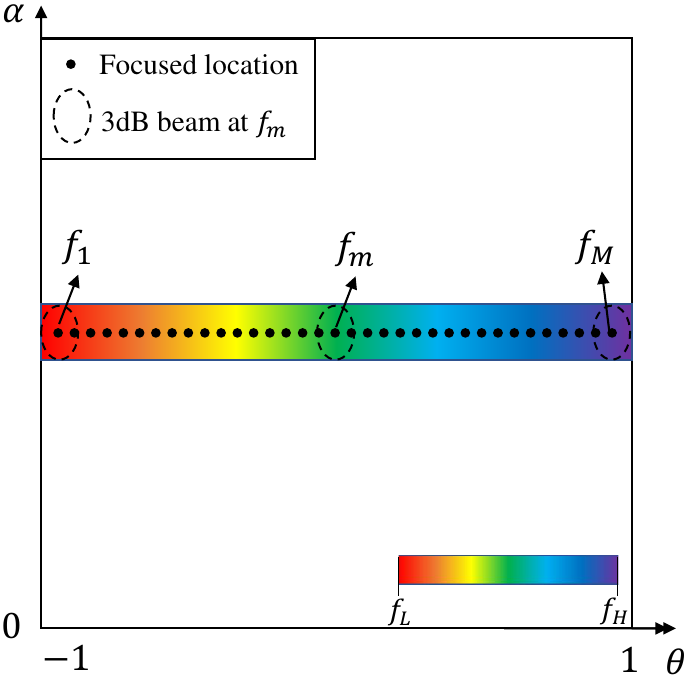}}
	\subfigure[]{
		\includegraphics[width=0.3 \linewidth]{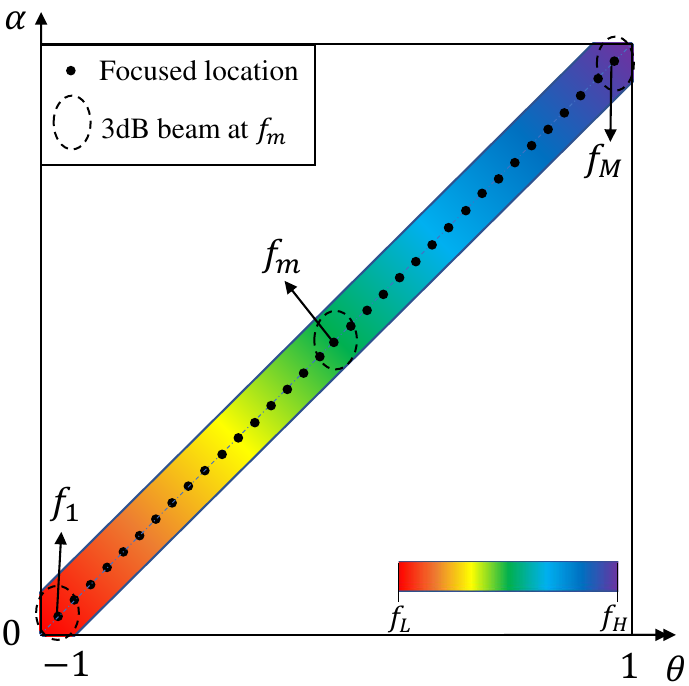}}
	\caption{Comparison of distance-dependent beam split with traditional beam split with different settings: (a) Beams of different subcarriers in proposed distance-dependent beam split; (b) Beams of different subcarriers in traditional beam split with $\alpha_p'+\frac{2q}{d}=0$; (c) Beams of different subcarriers in traditional beam split with $\alpha_p'+\frac{2q}{d}\neq 0$.}
	\label{pic_skip}
\end{figure*}
\else 
\begin{figure}
	\centering 
	\subfigure[Beams of different subcarriers in proposed distance-dependent beam split.]{
		\includegraphics[width=0.45 \linewidth]{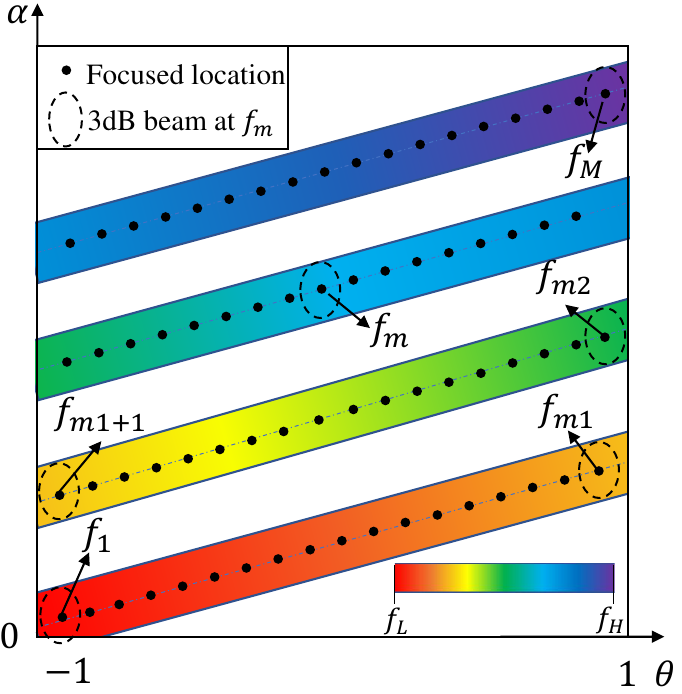}} 
	\subfigure[Beams of different subcarriers in linear beam split with $\alpha_p'+\frac{2q}{d}=0$.]{
		\includegraphics[width=0.45 \linewidth]{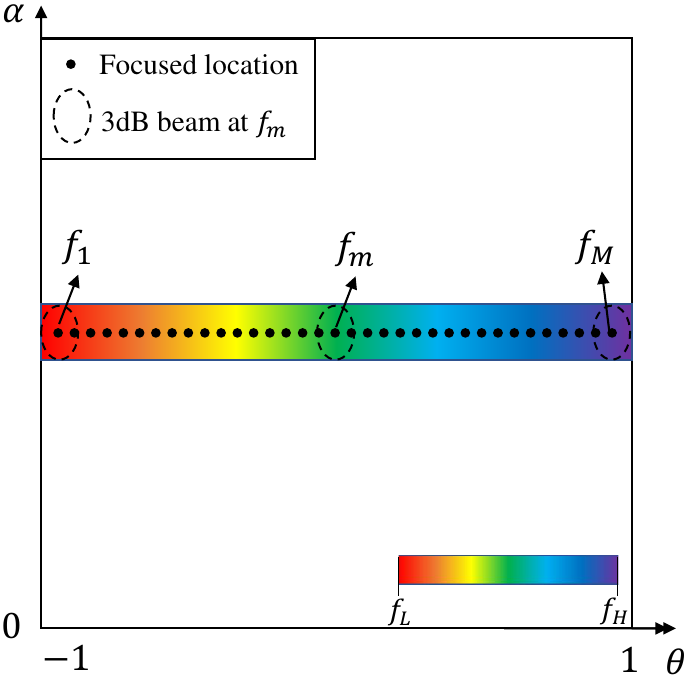}}
	\subfigure[Beams of different subcarriers in linear beam split with $\alpha_p'+\frac{2q}{d}\neq 0$.]{
		\includegraphics[width=0.45 \linewidth]{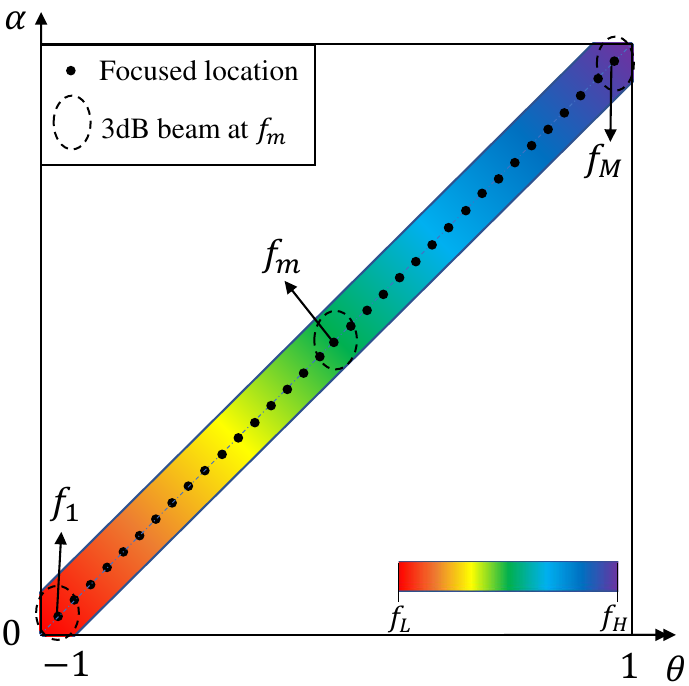}}
	\caption{Comparison of distance-dependent beam split with traditional linear beam split with different settings.}
	\label{pic_skip}
\end{figure}
\fi

{

\textbf{Distance-dependent beam split versus traditional near-field beam split}: To create the multi-strip distribution of beam split, it is necessary to make $\theta_p' + 2p$ large enough for triggering the rapid fluctuation in the angular domain and make  $\alpha_p'+\frac{2q}{d}$ non-zero for changing the distance ring monotonously. These requirements differ the distance-dependent beam split from the traditional counterpart in~\cite{nearrainbow}, which are compared in Fig.~\ref{pic_skip}.  
In~\cite{nearrainbow}, by setting $\theta_p'+2p$ with a \emph{non-zero but relatively small} value, the focused direction of the beams is linearly related to the frequency $f_m$ in \emph{a single period}; while by setting $\alpha_p'+\frac{2q}{d}=0$, the focused distance ring remains unchanged. Thus, beams of different frequencies focus in different angles in a single distance ring $\alpha= \alpha_t'$, or equivalently the single horizontal strip in Fig.~\ref{pic_skip}(b). Additionally, in Fig.~\ref{pic_skip}(c), we make $\alpha_p'+\frac{2q}{d}$ non-zero but keep the same $\theta_p'+2p$ as that of Fig.~\ref{pic_skip}(b). This beam-split pattern suggests that both the focused direction and distance ring increase monotonically with the frequency $f_m$.
However, without exploiting the periodicity of the angle-dimension beam split, the beams can only occupy one inclined strip while the multi-strip distribution is still concealed. 
}


\subsection{Distance-Dependent Beam Split based Near-field Beam Training} \label{Sec_3_Subsec_2}
{
Applying the discovered distance-dependent beam split phenomenon, it is promising that during beam training, different angles and distances can be searched simultaneously using one pilot. Thereby, a few pilots are capable of covering the whole two-dimensional space. This observation motivates us to design a new near-field beam training method based on this phenomenon.  
The design principles, the configuration of TD-PS parameters, and detailed beam training procedures are presented one by one in the sequel.

\ifx\onecol\undefined
\begin{figure}
	\centering 
	\includegraphics[width=0.7\linewidth]{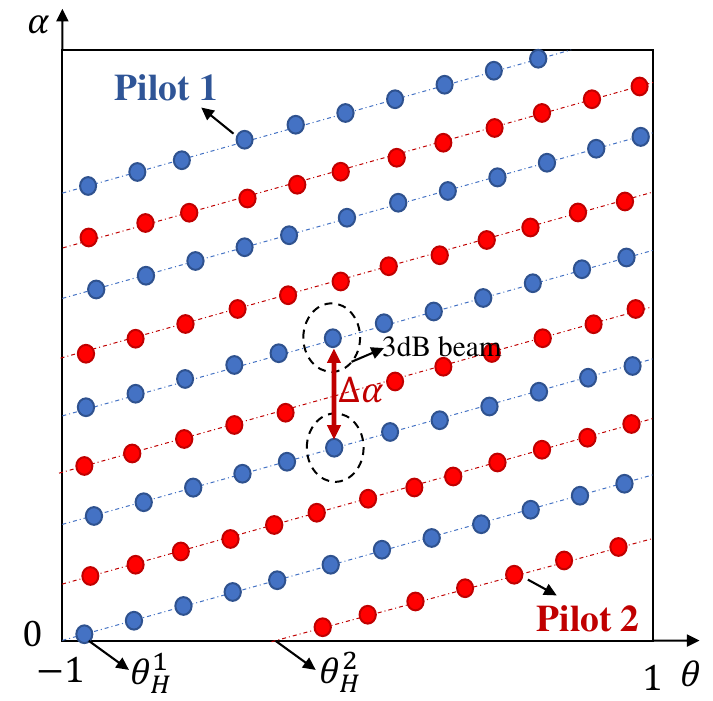}
	\caption{Add another pilot to guarantee coverage of possible regions.}
	\label{fig:two}
\end{figure}
\else 
\begin{figure}
	\centering 
	\includegraphics[width=0.45\linewidth]{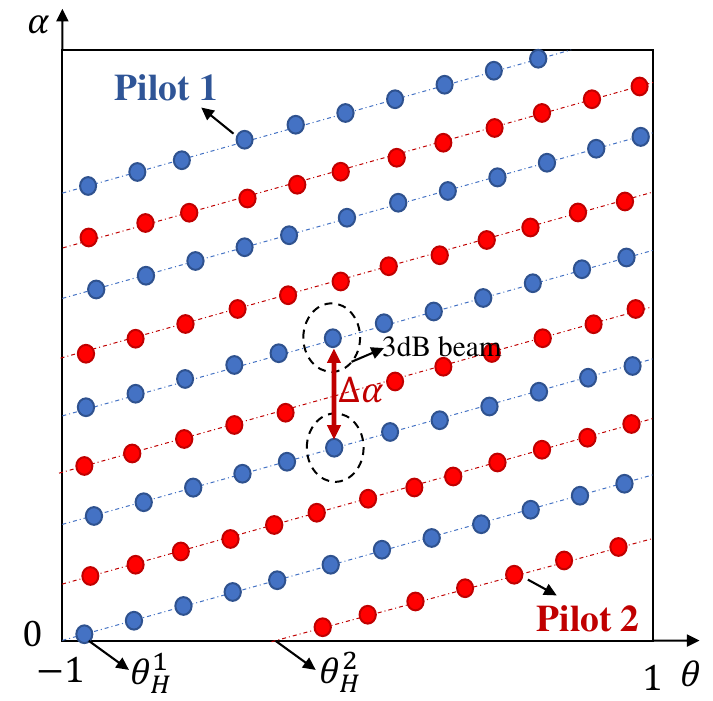}
	\caption{Add another pilot to guarantee coverage of possible regions.}
	\label{fig:two}
\end{figure}
\fi

\subsubsection{Basic design principles}
The basic principles for designing near-field beam training aided by distance-dependent beam split involve the consideration of reliability and efficiency.
\begin{itemize}
	\item {\bf Reliability.} To support reliable beam training, each location inside the communication region must be covered by one training beam. To quantify this metric, we require that each location within the ``angle-distance'' region lies in the $3~{\rm dB}$-beam width of a certain beam. 
	\item {\bf Efficiency.} In addition to reliability, we aim to use as few pilots as possible to improve training efficiency.
\end{itemize}
}


\subsubsection{Configuration of TD-PS parameters}
Building on the above principles, we elaborate on the manipulation of TD-PS parameters to create multi-strip beam pattern for beam training. 
We start from the manipulation of a single pilot to guarantee the  $3~{\rm dB}$ coverage in the angle domain. Then we will explain how to utilize interleaved pilots to guarantee the  $3~{\rm dB}$ coverage in the distance domain.

For the first pilot, let's consider the angle-related parameters $\theta_t'$ and $\theta_p'$, as well as the integer $p_m$ first. 
Based on~\eqref{eq_change}, the focused direction difference of adjacent subcarriers on the same strip with $p_m=p_{m+1}$ is
\begin{equation}\label{eq:dif}
    |\theta_{m+1}-\theta_m| = |f_c (\theta_p'+2p_m) (\frac{1}{f_{m+1}}-\frac{1}{f_m})|.
\end{equation}
For ease of discussion, we assume $\theta_p'+2p_m$ is always greater than 0 in the sequel, while all our conclusions can be straightforwardly generalized to negative $\theta_p'+2p_m$ situations. 
Formula~\eqref{eq:dif} suggests that a larger $\theta_p'+2p_m$ leads to a faster change in the focused direction versus frequency $f_m$. Thus, a relatively large $\theta_p'+2p_m$ can broaden the coverage area of beams, thereby improving beam training efficiency. 
However, since the direction difference in \eqref{eq:dif} grows linearly with  $\theta_p'+2p_m$, excessively large $\theta_p'+2p_m$ will lead to uncovered angular range between adjacent subcarriers. Taking these two factors into consideration, \textbf{Lemma 1} provides an upper limit of $\theta_p'+2p_m$ that can avoid angular coverage hole.
\begin{lemma} \label{lem1_angle}
Consider an arbitrary physical location $(\bar{\theta}, \bar{\alpha})$. 
If $\theta_p'+2p_m >0$, to make the direction $\bar{\theta}$ lie in the angle-domain $3 ~ {\rm dB}$-beamwidth of a certain beam, $\theta_p'$ should satisfy
	\begin{equation} \label{eq_t2}
		\theta_p' \in \left\{\theta_p': \theta_p'+2p_m \leqslant \frac{0.88}{N_t}(f_m+f_{m+1})\frac{M}{B} , \forall p_m\in\mathbb{Z}\right\}.
	\end{equation} 
\end{lemma}
\begin{IEEEproof}
(See Appendix A).
\end{IEEEproof}
It is easy to prove that when $\theta_p'+2p_m >0$, $p_m$ is a non-decreasing function with frequency $f_m$ (i.e. $\theta_p'+2p_m \leqslant \theta_p'+2p_M $). Besides, $ \frac{0.88}{N_t}(f_m+f_{m+1})\frac{M}{B}$ has minimum value at $f_m=f_L$.
Thus, by setting
\begin{equation}\label{eq_t3}
    \theta_p' \in \left\{\theta_p': \theta_p'+2p_M \leqslant \frac{1.76f_L M}{N_t B}, \forall p_M\in\mathbb{Z}\right\},
\end{equation}
\eqref{eq_t2} holds for all subcarriers. 
Therefore, by setting $p_M = \lfloor  \frac{0.88 \gamma f_L M}{N_t B} \rfloor$, $\theta_p'$ can be determined by 
\begin{equation}\label{eq_t4}
    \theta_p' = \frac{1.76\gamma f_L M}{N_t B} -2p_M,
\end{equation}
where $\gamma \in [0,1]$ is a hyperparameter. When $\gamma=1$, the equality in~\eqref{eq_t3} holds, while otherwise, it denotes an oversampling strategy in the angle domain.

While the PS parameter $\theta_p'$ influences the direction difference, the TD parameter $\theta_t'$ determines the intercept of the strip according to~\eqref{eq_appro}. 
In other words, it affects the starting/ending direction (i.e. $\theta_1$ or $\theta_M$) of this pilot. Without loss of generality, we expect that the ending direction of the beam at frequency $f_M$ is $\theta_M=1$. Accordingly, the parameter $\theta_t'$ can be calculated by
\begin{equation}\label{eq_theta1}
	\theta_t' = 1-\frac{f_c}{f_M}(\theta_p'+2p_M).
\end{equation} 
For future usage, here we also calculate the integer $p_1$ of the first subcarrier. $p_1$ should be automatically adjusted to satisfy the angle constraint $\theta_t'+\frac{f_c}{f_L}(\theta_p'+2p_1) \in [-1,1]$, so it can be presented by
\begin{equation} \label{eq:p1}
    p_1 = {\rm round} \left ( \frac{-\theta_t' \frac{f_L}{f_c}- \theta_p'}{2} \right).
\end{equation}

Next, we continue to discuss the manipulation of distance-related parameters $\alpha_t'$ and $\alpha_p'$, and the corresponding integer $q$. To guarantee that beams can cover the desired distance range, $\alpha_t'$ and $\alpha_p'$ should meet the requirements in~\eqref{eq_alphaL} and ~\eqref{eq_alphaH}.  We assume that the possible user distribution distance is $[\alpha_{\rm min},\alpha_{\rm max}]$. By setting $(\alpha_{p}, \alpha_{q})= (\alpha_{\rm min}, \alpha_{\rm max})$ and subtracting \eqref{eq_alphaH} from \eqref{eq_alphaL}, $\alpha_p'+\frac{2q}{d}$ is supposed to satisfy
\begin{equation}\label{eq_alpha2}
        \alpha_p' \in \left\{\alpha_p': \alpha_p'+\frac{2q}{d} 
        \ge 
        \frac{ \alpha_{\rm max}- \alpha_{\rm min}}{\frac{f_c}{f_L}-\frac{f_c}{f_H}}, \forall q \in\mathbb{Z}\right\}.
\end{equation}
Therefore, by setting $q = \lfloor \frac{ \alpha_{\rm max}- \alpha_{\rm min}}{\frac{f_c}{f_L}-\frac{f_c}{f_H}} \frac{d}{2}\rfloor$, $\alpha_p'$ can be set as
\begin{equation}\label{eq_alpha3}
    \alpha_p' =  \frac{ \alpha_{\rm max}- \alpha_{\rm min}}{\frac{f_c}{f_L}-\frac{f_c}{f_H}} - \frac{2q}{d},
\end{equation}
After obtaining $\alpha_p'$, the parameter $\alpha_t'$ can be arbitrarily selected from the range 
\begin{align}\label{eq_alpha4}
\alpha_t' \in [\alpha_{\rm max}-\frac{f_c}{f_L} (\alpha_p'+\frac{2q}{d}), \alpha_{\rm min}-\frac{f_c}{f_H}(\alpha_p'+\frac{2q}{d})], 
\end{align}
to meet the constraints \eqref{eq_alphaL} and \eqref{eq_alphaH}.

\begin{remark}
In summary, the parameters designed in \eqref{eq_t4}, \eqref{eq_theta1}, \eqref{eq_alpha3}, and \eqref{eq_alpha4} allow us to create the multi-strip beam pattern in Fig.~\ref{pic_skip}(a) using one single pilot. These beams can guarantee the 3 dB coverage of the entire angular range in each strip.
\end{remark}

However, one remaining issue is that the $3~{\rm dB}$ coverage in the distance domain may not be satisfied, since there exists coverage hole between adjacent strips as shown in Fig.~\ref{pic_skip}(a). 
Specifically, consider two searched locations at the same angle but adjacent strips as illustrated by the circled blue points in Fig.~\ref{fig:two}. 
Based on \eqref{eq_appro} and \eqref{eq_alpha_appro}, the angle and distance differences, $\Delta \theta$ and $\Delta \alpha$, of these two locations can be calculated as 
\begin{align}
    \Delta  \theta &= \frac{\theta_p'+2p_m}{f_c} \Delta f =2, \label{eq:delta_theta}\\
    \Delta  \alpha &= \frac{\alpha_p'+\frac{2q}{d}}{f_c} \Delta f, \label{eq:delta_alpha}
\end{align}
where $\Delta f$ denotes the frequency difference. From equations \eqref{eq:delta_theta} and \eqref{eq:delta_alpha}, the distance difference $\Delta \alpha $ can be expressed as
\begin{equation}
	\Delta \alpha = \frac{2(\alpha_p'+\frac{2q}{d})}{\theta_p'+2p_m} \leqslant \frac{2(\alpha_p'+\frac{2q}{d})}{\theta_p'+2p_1} \overset{\Delta}{=} \Delta \alpha^{\rm upper}.
\end{equation}
To realize the coverage over the entire distance domain, we expect that the distance difference $\Delta \alpha$ is smaller than the sum of the distance-domain $3 ~ {\rm dB}$-beam width of the two beams, which can be presented as 
\begin{equation}\label{eq_angle}
	\Delta \alpha \leqslant \eta_\alpha^{f_a}+ \eta_\alpha^{f_b}.
\end{equation}
Here, $f_a$ and $f_b$ refer to the frequency of the corresponding beams and $\eta_\alpha^{f}$ is the distance-domain $3 ~ {\rm dB}$-beam width at frequency $f$. As proven in Appendix B, the beam width $\eta_\alpha^{f}$ has an analytical expression: $\eta_\alpha^{f} = \frac{4\beta^2 f_c^2}{N_t^2 c f}$, where $\beta = 1.318$.

In practice, however, the constraint~\eqref{eq_angle} is hard to satisfy, indicating that a single pilot is not sufficient to cover all distances. 
To fill up the coverage hole of a single pilot, we plug in more pilots to create \emph{interleaved multi-strip beam pattern}, as illustrated in Fig.~\ref{fig:two}. 
To this aim, recall that the intercept of a strip is dependent on the TD parameter $\theta_t'$ according to~\eqref{eq_appro}. By changing the value of $\theta_t'$, the ending direction $\theta_M$ is modified and thus the strips of one pilot are shifted. This observation motivates us to retain the parameters $\theta_p', \alpha_t'$, and $\alpha_p'$ but set  $\theta_t'$ as different values for different pilots to create the interleaved multi-strip beam pattern. We disperse different pilots to create uniformly spaced strips (the circled red points in Fig.~\ref{fig:two}) to efficiently fill in the uncovered ranges.

To elaborate, we denote the total pilot number as $K$ and introduce the superscript $k$ to index the TD parameter $\theta_t'^k$ and the ending direction $\theta_M^k$ of the $k$-th pilot. The minimal pilot number required for distance coverage is given as follows.  
\begin{remark} 
The $K$ pilots need to fill up the distance-domain coverage hole with an upper length of $\Delta \alpha^{\rm upper}$. Given that the distance-domain beam width has a lower bound $\eta^f_\alpha \ge \eta^{f_H}_\alpha$, 
the minimal pilot number is determined by 
\begin{align}\label{eq_num}
    K \ge \frac{\Delta \alpha^{\rm upper}}{2 \eta^{f_H}_\alpha} = \frac{(\alpha_p'+\frac{2q}{d}) N_t^2 c f_H  }{4(\theta_p'+2p_1)\beta^2 f_c^2}.
\end{align}
\end{remark}
After obtaining the pilot number $K$, we can uniformly sample the ending directions from the range $[-1, 1]$ such that $\theta_M^k = 1 - \frac{2}{K}(k-1)$ for $k \in\{1,2,\cdots, K\}$ to ensure uniformly spaced strips. Therefore, $\{\theta_t'^k\}$ for each pilot are given as 
\begin{equation}\label{eq_theta10}
	\theta_t'^k = \theta_M^k - \frac{f_c}{f_M}(\theta_p'+2p_M), \quad k\in\{1,2\cdots, K\}.
\end{equation}
It is notable that the parameter $\theta_t'$ designed in \eqref{eq_theta1} is exactly the first parameter $\theta_t'^1$ designed in \eqref{eq_theta10} since $\theta_M^1 = 1$. 

Based on the discussions above, we summarize the TD-PS parameter manipulation procedures in {\bf Algorithm \ref{alg:mani}}. For Step 1-3, by setting the angle-related parameters of the first pilot, we manipulate the beams to periodically cover the angular range while ensuring the $3$-dB coverage in the angle domain. Then based on Step 4, the beams are set to cover the desired distance range. Finally, we utilize interleaved pilots to ensure $3$-dB coverage in distance domain, as shown in Step 5-6.
\begin{algorithm}[htb]
	\caption{ TD-PS parameter manipulation for distance-dependent beam-split-based beam training.}
	\label{alg:mani}
	\renewcommand{\algorithmicrequire}{\textbf{Input:}}
	\renewcommand{\algorithmicensure}{\textbf{Output:}}
	\begin{algorithmic}[1] 
	\REQUIRE $N_t$, $f_c$, $B$, $M$, $\alpha_{\rm max}$, $\alpha_{\rm min}$, and $\gamma$.\\
	\STATE   Set $p_M = \lfloor  \frac{0.88 \gamma f_L M}{N_t B} \rfloor$ and obtain $ \theta_p'$ by \eqref{eq_t4}; 
    \STATE   Set $\theta_M^1 = 1$ and obtain $ \theta_t'^1$ for the first pilot by \eqref{eq_theta1}; 
    \STATE Set $p_1$ as \eqref{eq:p1};
	\STATE   Set $q = \lfloor \frac{ \alpha_{\rm max}- \alpha_{\rm min}}{\frac{f_c}{f_L}-\frac{f_c}{f_H}} \frac{d}{2}\rfloor$ and obtain $ \alpha_p'$ and $\alpha_t'$ based on \eqref{eq_alpha3} and \eqref{eq_alpha4};
	\STATE   Calculate the required pilot number $K$ based on \eqref{eq_num};
	\STATE   Obtain $\{\theta_t'^k\}$ for the remaining $K-1$ pilots based on \eqref{eq_theta10}.

	\ENSURE Designed parameters $\{\theta_t'^k\}$, $\theta_p'$, $ \alpha_t'$ and $ \alpha_p'$.
	\end{algorithmic}
\end{algorithm}

\subsubsection{Beam training process}
In the beam training process, the BS sequentially transmits $K$ pilots to the user. The received array gains over all subcarriers and pilots are written as
\begin{equation}
	\begin{split}
		{\bm g} = &[g(1,1), g(2,1), \cdots, g(M,1), \\
		 &\cdots, g(1,K),  g(2,K), \cdots, g(M,K)],
	\end{split}
\end{equation}
where $g(m,k)$ is the array gain at the $m$-th subcarrier of the $k$-th pilot. The user selects the subcarrier with the highest received power (i.e. the strongest element of ${\bm g}(\hat{m},\hat{k})$) and the estimated physical location can be calculated based on~\eqref{eq_change} and \eqref{eq_alpha}.

\subsection{Hardware Implementations of Proposed Method}
This sub-section
introduces the hardware implementation of the TD-PS structure to realize the proposed beam training algorithm. In particular, we mainly focus on the expensive TD circuits. 
The function of \emph{time delay} can be implemented mainly by two kinds of approaches, namely the adjustable true-time-delay-based methods~\cite{tdhard1,tdhard2} and fixed true-time-delay-based methods~\cite{tdfixed,tddrawback}. The former one can provide continuously adjustable time delay over the entire range with significantly high power consumption and hardware complexity~\cite{tddrawback}. In contrast, the available time-delays of fixed true-time-delay-based method are restricted to several fixed values. Thus, this architecture has much lower power consumption and hardware complexity, with the sacrifice of TD resolution.


\ifx\onecol\undefined
\begin{figure}
	\centering 
		\includegraphics[width= 0.85 \linewidth]{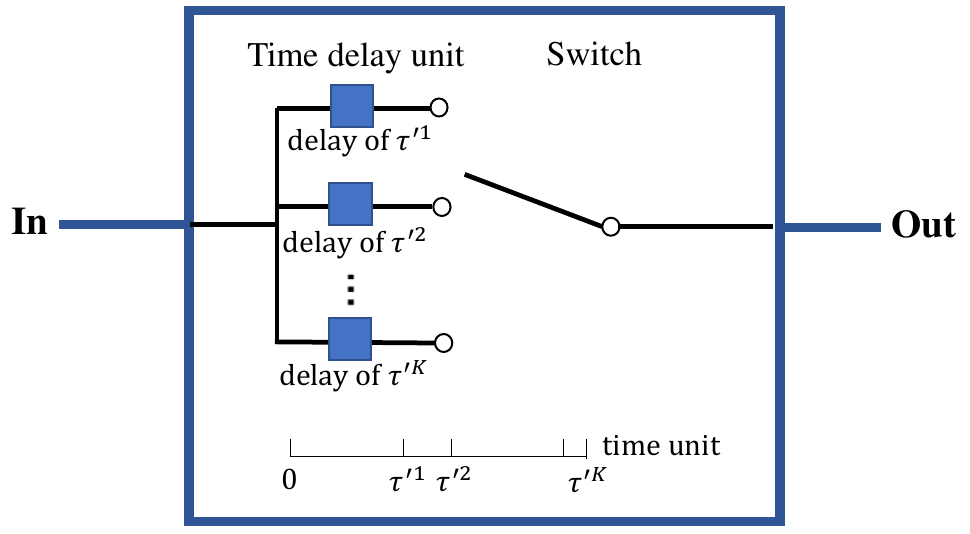}
	\caption{Hardware implementations of TD by fixed time delay network.}
	\label{pic_imple}
\end{figure}
\else 
\begin{figure}
	\subfigure[Adjustable time delay.]{
		\begin{minipage}[c]{0.5\textwidth}
		  \centering
		  \includegraphics[width=\textwidth]{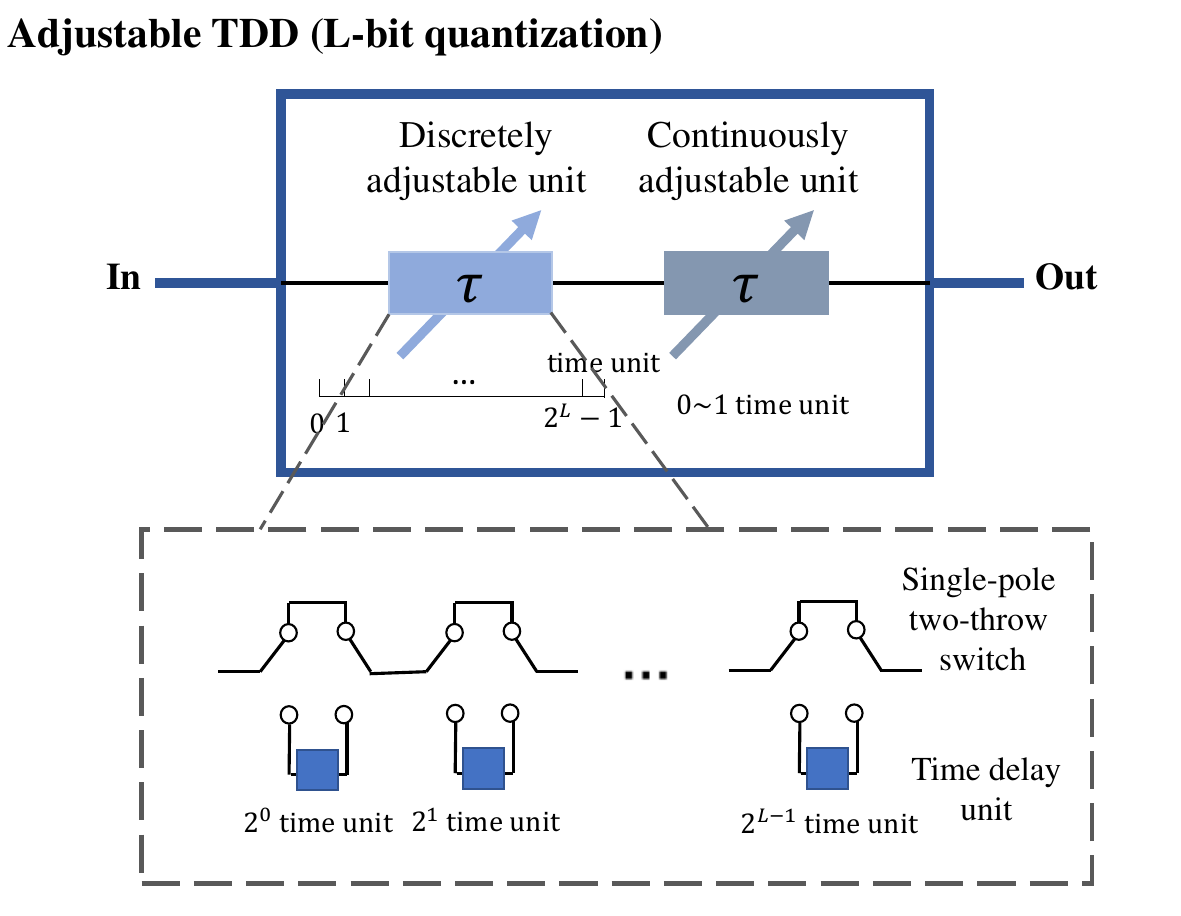}
		\end{minipage}
	  }
	\subfigure[Fixed time delay.]{
		\begin{minipage}[c]{0.4\textwidth}
		  \centering
		  \includegraphics[width=\textwidth]{figures/FixedTD.pdf}
		\end{minipage}
	  }
	\caption{Typical hardware implementations of adjustable TD and fixed TD.}
	\label{pic_imple}
\end{figure}
\fi



Fortunately, the low-cost fixed true-time-delay-based methods could adequately satisfy the requirements for realizing the proposed beam training algorithm.
To be specific,
the scanned positions of beams are fixed and the corresponding time delay values are elaborately designed and can be determined and calculated in advance. Besides, 
only $K$ (usually a small number such as 2,3) time delay values need to be realized by TD. Therefore, as shown in Fig.~\ref{pic_imple},
we propose to deploy a nonuniform-quantization TD with fixed time delay unit and a switch network, which can be implemented by an extensively-utilized microstrip line or waveguide of fixed length~\cite{tdfixed}.
For the $n$-th TD, we calculate the $K$ desired time delay value as $\tau^{(n)'k},k=1,2,\cdots,K$: 
\begin{equation}\label{time}
	\tau^{(n)'} = \frac{r_1^{(n)'}}{c}=\frac{nd\theta_t'-n^2d^2\alpha_t'}{c}.
\end{equation}
Then we construct the TD with the fixed TD comprising $K$ fixed time delay units and a switch, indicating the selected fixed TD in different time slots. In this context, only $\log_2 K$ bits are required to represent the selected time delay.

Note that the incorporation of TDs may introduce additional Insertion Loss (IL) into the link budget, indicating the signal attenuation experienced as it traverses the hardware device. For the proposed implementation, the IL associated with the TD is the cumulative effect of the insertion losses from one fixed time delay unit and one switch, i.e., $IL_{FTD} +IL_{switch}$, which is significantly lower than adjustable TDs~\cite{tdfixed}.
Besides, as discussed in~\cite{wuucawide}, additional IL introduced by fixed TD may not have significant influence on beamforming gain and it can be compensated by improving the amplification factor of the power amplifiers. 


\section{Improved Beam Training for Off-grid Users} \label{offgrid}
Compared to the method in~\cite{nearrainbow}, the sampled density of direction within $[-1,1]$ of the proposed method in Section~\ref{proposed} is relatively low, which limits its performance for off-grid users (i.e. users not on the searched locations). 
In this section, we propose two algorithms, namely the auxiliary beam pair-assisted method and the match filter-based method, to solve this problem.
The core idea of the proposed methods is to employ the received power of multiple subcarriers, rather than only the one with the highest received power to jointly estimate the users' locations.

\subsection{Auxiliary beam pair-assisted method}
\ifx\onecol\undefined
\begin{figure}
	\centering 
	\includegraphics[width=0.75 \linewidth]{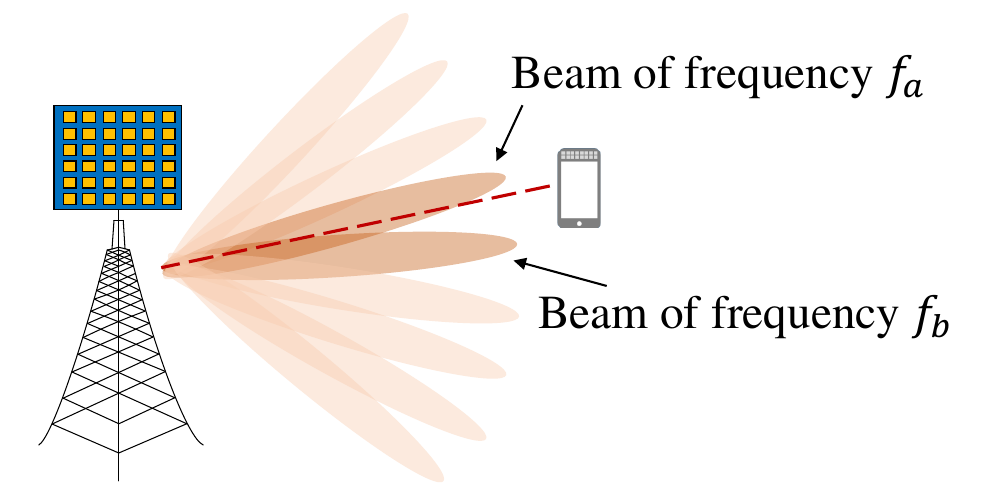}
	\caption{Auxiliary beam pair-assisted method.}
	\label{fig:aux}
\end{figure}
\else 
\begin{figure}
	\centering 
	\includegraphics[width=0.45\linewidth]{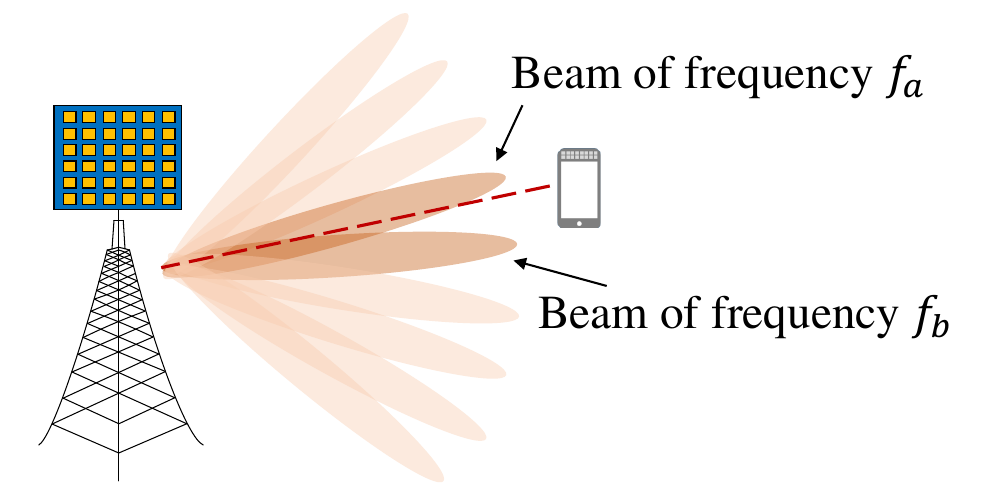}
	\caption{Auxiliary beam pair-assisted method.}
	\label{fig:aux}
\end{figure}
\fi

As shown in Fig.~\ref{fig:aux}, when the user is located between two beams of frequency $f_a$ and $f_b$, the scheme in Section~\ref{proposed}, which chooses the codeword having the highest received power at frequency $f_a$, is suboptimal. Inspired by~\cite{auxibeam,auxibeam2}, we design an auxiliary beam pair-assisted beam training method based on near-field distance-dependent beam split. The method utilizes the amplitude distribution of the auxiliary beam pair, $f_a$ and $f_b$, to accurately estimate the angle and distance of off-grid users. To this aim, we will first analyze the amplitude distribution of near-field beams, and then the beam training method is elaborated.

\subsubsection{Analysis of amplitude distribution of near-field beams}
The equal-power contour of near-field beams can be clarified by the following {\bf Proposition 1}. 
\begin{proposition}\label{pro1_distance}
	The equal-power contour of near-field beams can be approximated as an ellipse equation:
 	\begin{equation} \label{eq_ell}
		G(k(\theta-\bar{\theta}),k(\alpha-\bar{\alpha}))  = 1- \sigma_1 (\theta-\bar{\theta})^2 - \sigma_2 (\alpha-\bar{\alpha})^2,
	\end{equation}
	where  $(\theta,\alpha)$ is the focused location of the beam, $(\bar{\theta},\bar{\alpha})$ is an arbitrary physical location within the curve of $3 ~ {\rm dB}$-beam width. Parameters $\sigma_1,\sigma_2$ are  presented as 
	\begin{align}
		\sigma_1 =\frac{1}{24}N_t^2 \pi^2 \frac{f^2}{f_c} \quad {\rm and}\quad 
		\sigma_2 =\frac{\pi^2}{90} \frac{N_t^4 d^4}{\lambda^2}.
	\end{align}
\end{proposition}
\begin{IEEEproof}
    (See Appendix C).
\end{IEEEproof}

\subsubsection{Procedures of auxiliary beam pair-assisted method}
According to the analysis above, the received array gain of the beam at frequency $f_a$ and $f_b$ each defines an ellipse in which the user is located. The angle and distance to be estimated can be obtained through the intersection of the ellipses. This is the basic idea of the auxiliary beam pair-assisted method. Then we will illustrate the procedures step by step as follows.
\begin{itemize}
	\item Firstly, we perform the beam training proposed in Section~\ref{proposed} to obtain subcarrier $f_a$ with the highest received power. Suppose we can obtain the value of the highest power $P_a$, we calculate the array gain as $G_a$. Besides, we
    denote the focus location of the beam as $(\theta_a,\alpha_a)$. 
	\item Then we compare the received power at subcarriers $f_{a-1}$ and $f_{a+1}$, and select the one with higher power, denoted as $f_{b}$. Similarly, record the focus location $(\theta_b,\alpha_b)$ and array gain as $G_b$. 
	\item The corresponding equal-amplitude ellipse defined by  subcarrier $f_a$ and $f_b$  can be written as 
	\begin{equation}
		\left\{\begin{array}{cc} \sigma_1 (\theta_a-\theta)^2 + \sigma_2 (\alpha_a-\alpha)^2 &= 1-G_a\\
			\sigma_1 (\theta_b-\theta)^2 + \sigma_2 (\alpha_b-\alpha)^2 &= 1-G_b
		\end{array} \right. , 
	\end{equation}
	based on which the estimated $\hat{\theta }$ and $\hat{\alpha  }$ can be obtained using the Newton method.
\end{itemize}

\vspace{-5mm}
\subsection{Match filter-based method}
The auxiliary beam pair-assisted beam training utilizes a pair of beams to improve the training performance for off-grid users. In this subsection, we generalize it to a match filter-based beam training method, which advocates the use of array gains over all subcarriers and pilots for accurate beam training. 


In this method, a substantial number of grids with different angles and distances are sampled offline. For each grid, the distribution of array gain over different subcarriers and pilots can be calculated by the location of the grid. Then, in the online stage, we select the grid whose array gain distribution matches the best with the received array gain distribution of the user. 
Since the sampled grids can be much denser than the searched locations, the performance for off-grid users is improved. Moreover, the characteristics of the match filter also protect training from noise. The framework of this method is elaborated below.
\begin{itemize}
	\item \textbf{Offline stage}: Sample the possible user locations by $L$ grids on the angle domain and $S$ grids on the distance domain. The sampled grids are supposed to be denser than the searched locations by the proposed beam training method.
 For the sampled location $(\theta_l,\alpha_s)$, the array gain at the $m$-th subcarrier of the $k$-th pilot is presented as
	\begin{align}
		&g_{l,s}(m,k) \notag \\
  &=|({\bf w}_m^{\rm TD}(\theta_t'^k,\alpha_t') \odot  {\bf w}^{\rm PS}(\theta_p',\alpha_p'))^T {\bf b}_m(\theta_l, \alpha_s)|.
	\end{align}
Then, the corresponding array gain vector, ${\bm g}_{l,s}$, is
	\begin{align}
			&{\bm g}_{l,s} = [g_{l,s}(1,1), g_{l,s}(2,1), \cdots, g_{l,s}(M,1), \notag\\ &\:\: \cdots, g_{l,s}(1,K),  g_{l,s}(2,K), \cdots, g_{l,s}(M,K)].
	\end{align}
	\item \textbf{Online stage}: For each of the $L \times S$ grids, the similarity between the array gain vector of location $(\theta_l,\alpha_s)$ and the received array gain vector of the user is calculated as
	\begin{equation}
		r_{l, s} = {\bm g}^T {\bm g}_{l, s}, s=1,2,\cdots,S, l=1,2,\cdots,L.
	\end{equation}
 Finally, we select the grid with largest similarity $r_{l, s}$ and the estimated location of the user is $\hat{\theta}=\theta_l$ and $\hat{\alpha}=\alpha_s$.
\end{itemize}
To reduce feedback overhead, the users are supposed to estimate $\hat{\theta }$ and $\hat{\alpha  }$ locally and feed back the results to the BS, rather than feeding back all the received power to the BS.

\section{Simulation Results} \label{sec-re}
In this section, we present simulations to validate the proposed distance-dependent beam split phenomenon. Subsequently, the performance of the proposed distance-dependent beam split-based beam training schemes, encompassing both on-grid algorithms and off-grid algorithms, is assessed.

\subsection{The Demonstration of Distance-dependent Beam Split}

\ifx\onecol\undefined
\begin{figure*}
	\centering 
	\subfigure[Distance-dependent beam split with a single pilot.]{
		\includegraphics[width= \linewidth]{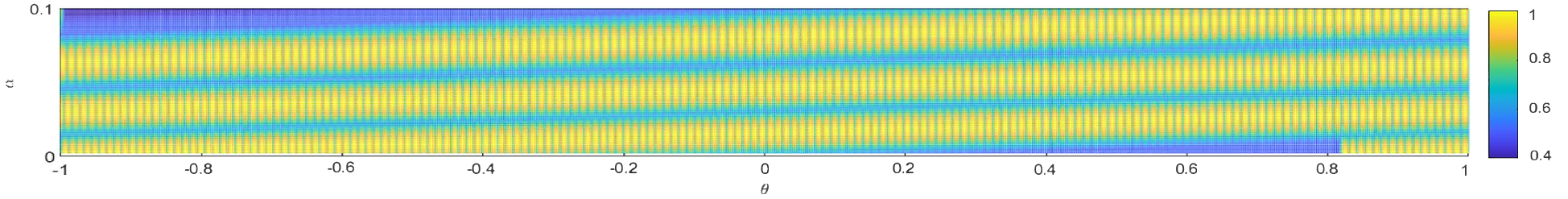}} 
	\subfigure[Distance-dependent beam split with two pilots.]{
		\includegraphics[width= \linewidth]{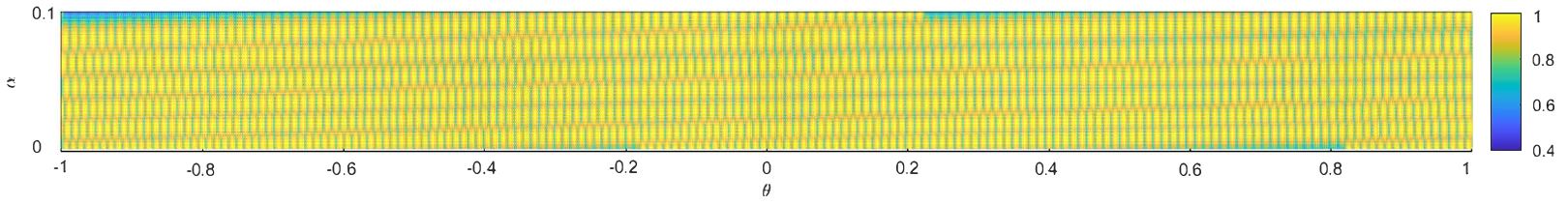}}
	\caption{Illustration of near field near-field distance-dependent beam split.}
	\label{pic_beam}
\end{figure*}
\else 
\begin{figure}
	\centering 
	\subfigure[Distance-dependent beam split with a single pilot.]{
		\includegraphics[width=0.9 \linewidth]{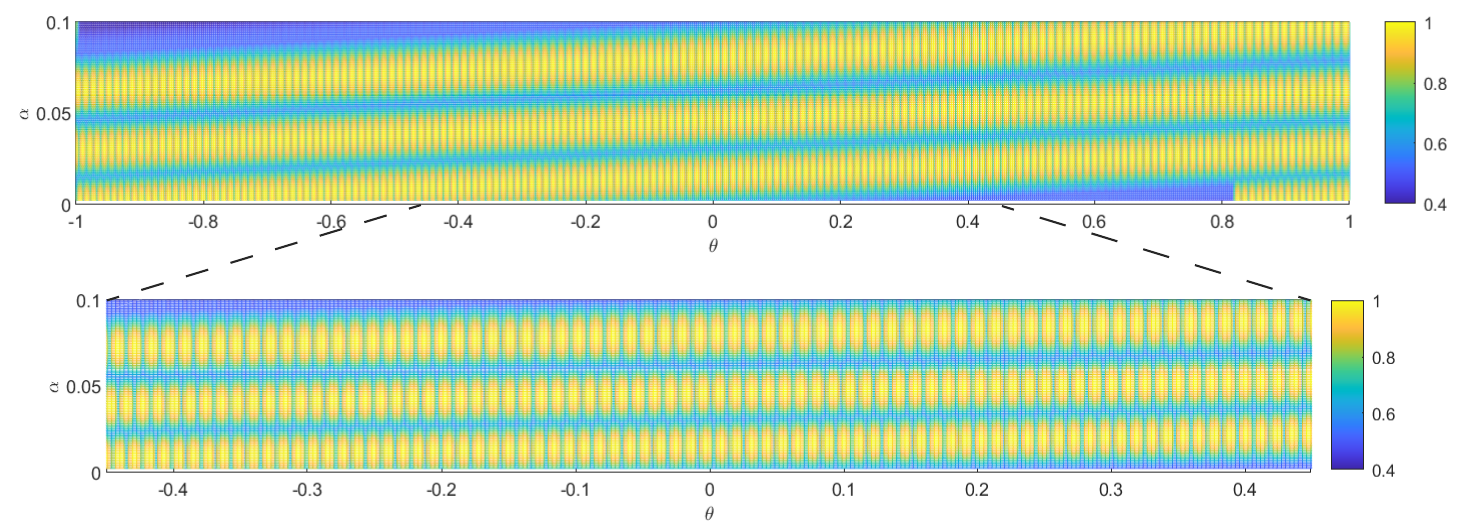}} 
	\subfigure[Distance-dependent beam split with two pilots.]{
		\includegraphics[width=0.9 \linewidth]{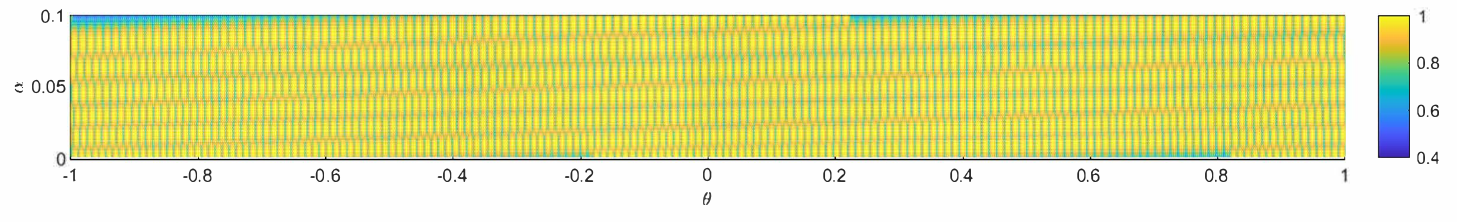}}
	\caption{Illustration of near field near-field distance-dependent beam split.}
	\label{pic_beam}
\end{figure}
\fi

We first verify that the designed TD-PS parameters in \textbf{Algorithm \ref{alg:mani}} are capable of producing distance-dependent near-field beam split.
As an example, we suppose the BS is equipped with $N_t=128$ antennas. The carrier frequency is $f_c =  10~{\rm GHz}$, the number of subcarriers is $M = 512 $, and the bandwidth is $B = 2 ~{\rm GHz}$. The users are uniformly distributed within angle range $[-\pi/3,\pi/3]$, and distance range $[\rho_{\rm min}, \rho_{\rm max}] = [5~{\rm m}, 200~{\rm m}]$, corresponding to a distance ring range $[\alpha_{\rm min}, \alpha_{\rm max}] = [1/400, 1/10]$. 
According to {\bf Algorithm \ref{alg:mani}}, the TD-PS parameters are defined as follows.
Based on~\eqref{eq_t3}, $\theta_p'+2p_M$ should satisfies $ \theta_p'+2p_M \leqslant \frac{1.76f_L}{N_t}\frac{M}{B} = 31.68$ 
and we set $\theta_p' = 1.68$, $p_M=15$ with $\gamma =1$. Then~\eqref{eq_theta1} 
and~\eqref{eq:p1} determine that $\theta_t'^1-27.8$ and $p_1=12$. 
Then based on~\eqref{eq_alpha2}, $\alpha_p'+\frac{2q}{d} \geqslant \frac{ \alpha_{\rm max}- \alpha_{\rm min}}{\frac{f_c}{f_L}-\frac{f_c}{f_H}}=0.483$, and  thus $\alpha_p'$ is set as $0.5$ with $q=0$ here.
Accordingly, $\alpha_t'$ is manipulated as $\alpha_t' = -0.454 \in [\alpha_{\rm max}-\frac{f_c}{f_L} (\alpha_p'+\frac{2q}{d}),\alpha_{\rm min}-\frac{f_c}{f_H} (\alpha_p'+\frac{2q}{d}) ] = [-0.456,-0.452]$.

We depict the distribution of beams over $M$ subcarriers generated by the TD-PS beamformer in Fig.~\ref{pic_beam} (a). 
As we expect, the focused points of beams create a multi-strip beam pattern, validating the effectiveness of distance-dependent beam split.
It is also observable that the beams of a single pilot in Fig.~\ref{pic_beam} (a) cannot cover the entire user region, leaving a coverage hole between strips. According to~\eqref{eq_num}, the minimum pilot number required is calculated as $K=2$. Thereby, we can add another pilot with the parameter $\theta_t'^2 = -28.8$ to make the second ending direction be $\theta_M^2 = 0$ based on~\eqref{eq_theta10}. The beam patterns of the two pilots are plotted together in Fig.~\ref{pic_beam} (b). 
In this context, the entire possible region is thoroughly searched, guaranteeing the reliability of beam training.

\subsection{Beam Training performance}
We then evaluate the performance of the proposed distance-dependent beam-split-based near-field beam training. The default simulation settings are as follows. 
The BS equips an $N_t = 256$-element ULA while the carrier frequency is $f_c = 30 ~{\rm GHz}$. The bandwidth is $B = 5~{\rm GHz}$ with subcarrier number $M = 1024$. 
The users are uniformly distributed within angle range $[-\pi/3,\pi/3]$, and distance range  $[\rho_{\rm min}, \rho_{\rm max}] = [5~{\rm m}, 200~{\rm m}]$, corresponding to a distance ring range   $[\alpha_{\rm min}, \alpha_{\rm max}] = [1/400, 1/10]$. 
According to \textbf{Algorithm \ref{alg:mani}}, we employ $K=3$ pilots for beam training. The TD-PS parameters are set as $\theta_p'=0.784$, $\{\theta_t'^k\}= \{ -32.95,-33.62,-34.29\}$, $\alpha_p' = 0.58$, and  $\alpha_t' = -0.53$, while the corresponding $\gamma = 0.95$, $p_1 = 15$, $p_M = 18$ and $q=0$.
To evaluate the effectiveness of beam training, we employ the average rate performance as a metric, which is presented as
\begin{equation}
	\begin{aligned}
		R & =\frac{1}{M} \sum_{m=1}^{M} \log _{2}\left(1+\frac{P_{t}}{\sigma^{2}}\left\|\mathbf{h}_{m}^{T} \mathbf{w}_{m}\right\|^{2}\right) \\
		& =\frac{1}{M} \sum_{m=1}^{M} \log _{2}\left(1+\frac{P_{t} N_{t} \beta_{m}^{2}}{\sigma^{2}}\left\|\mathbf{b}_{m}^{T}\left(\theta_{0}, r_{0}\right) \mathbf{w}_{m}\right\|^{2}\right)
	\end{aligned}		
\end{equation}
where ${\rm SNR} = \frac{P_{t} N_{t} \beta_{m}^{2}}{\sigma^{2}} $. The simulation results below are obtained by  $1000$ realizations of Monte Carlo trails.

\ifx\onecol\undefined
\begin{figure}
	\centering 
	\includegraphics[width=0.95\linewidth]{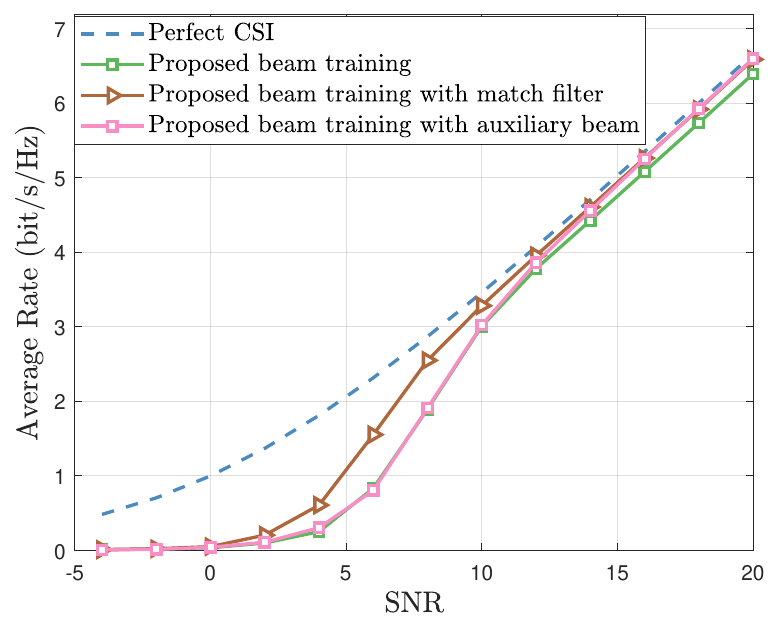}
	\caption{Average rate performance vs. SNR.}
	\label{fig:snr}
\end{figure}
\else 
\begin{figure}
	\centering 
	\includegraphics[width=0.5\linewidth]{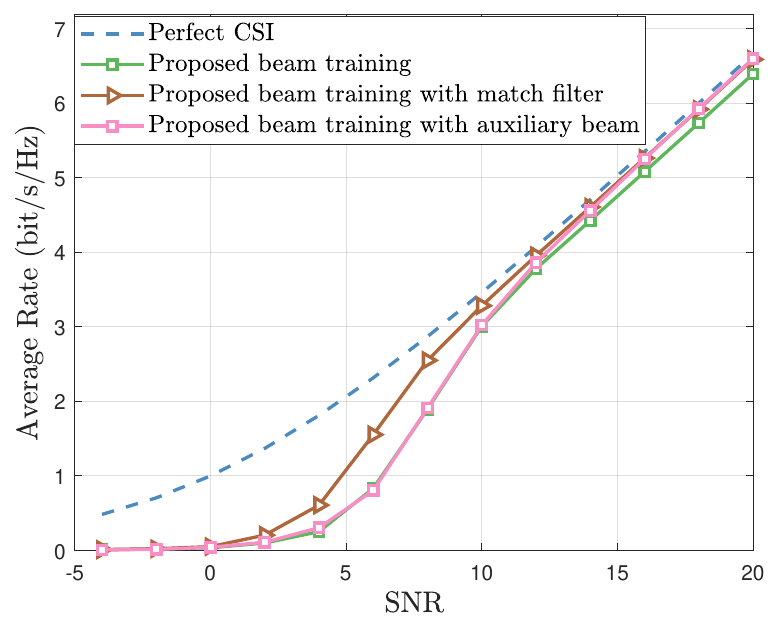}
	\caption{Average rate performance vs. SNR.}
	\label{fig:snr}
\end{figure}
\fi

To begin with, Fig.~\ref{fig:snr} presents the average rate performance against SNR ranging from $5~{\rm dB}$ to $20~{\rm dB}$. 
The proposed beam training algorithms for both on-grid and off-grid cases are included. 
The training overhead is $K=3$. The sampled grids in the angle and distance domain are $L=1024$ and $S=10$.
When the SNR is relatively high (e.g. ${\rm SNR}> 10~{\rm dB}$), both the proposed off-grid algorithms can improve the performance of distance-dependent beam split-based beam training performance. While at low SNR scenarios, the performance of the auxiliary beam pair-assisted method is almost the same as the on-grid method. This may result from the fact that an inaccurate beam gains $G_a$ and $G_b$ with noise poses the “error accumulation” phenomena when calculating the estimated $\hat{\theta }$ and $\hat{\alpha}$. Therefore, the auxiliary beam pair-assisted method might be vulnerable to noise. In contrast, thanks to the capability of restraining noise interference, match-filter-based method can ensure significant performance improvement for off-grid users even in low SNR regimes.

Then we compare the proposed distance-dependent beam-split-based near-field beam training enhanced by match filter with the following benchmarks:
\begin{itemize}
	\item {\bf Perfect CSI:}  BS has access to the perfect channel vector $\mathbf{h}_{m}$, which establishes the upper bound for performance.
	\item {\bf Near-field exhaustive beam training:} 
    This method conducts an exhaustive search across all potential directions and distances to identify the optimal codeword.
	\item {\bf Near-field rainbow based beam training:} As mentioned above, the near-field rainbow-based method~\cite{nearrainbow} searches multiple angles in a single distance ring with different subcarriers concurrently and sequentially searches through various distance rings exhaustively.
	\item {\bf Far-field rainbow-based beam training:} Proposed in~\cite{farsplit2}, this method employs TD beamforming to concurrently search for the optimal angle across different subcarriers while disregarding the distance information.
\end{itemize}

\begin{table*}[htb]
	\centering
	\caption{Comparisons of overheads for different Schemes}
	\begin{tabular}{ccc}
		\toprule
		Schemes & Training Overheads & Values \\
		\midrule
		Near-field exhaustive beam training & $TS$ & $10240$ \\
		Near-field rainbow based beam training & $S$ & $10$ \\
		Far-field rainbow  based beam training & $1$ & $1$ \\
		Proposed distance-dependent beam split based beam training & $K$ & $3$ \\
		\bottomrule
	\end{tabular}
	\vspace{0.5em}
	\label{tab:com}
	\vspace{-2em}
\end{table*}

Table~\ref{tab:com} summarizes the required pilot overheads for different techniques. 
We sample $L = 1024$ grids for the angle domain and $S = 10$ grids for the distance ring. Therefore, the pilot overheads of the near-field rainbow~\cite{nearrainbow} are $S = 10$, while for the near-field exhaustive method, the training overhead is up to $LS=10240$.
Besides, the pilot number is $K=3$ for proposed distance-dependent beam split-based beam training. 
As for the far-field rainbow technique, the optimal angle can be obtained with only a single pilot through TD beamforming, while the distance information is neglected.

\ifx\onecol\undefined
\begin{figure}
	\centering 
	\subfigure[]{
		\includegraphics[width=0.95\linewidth]{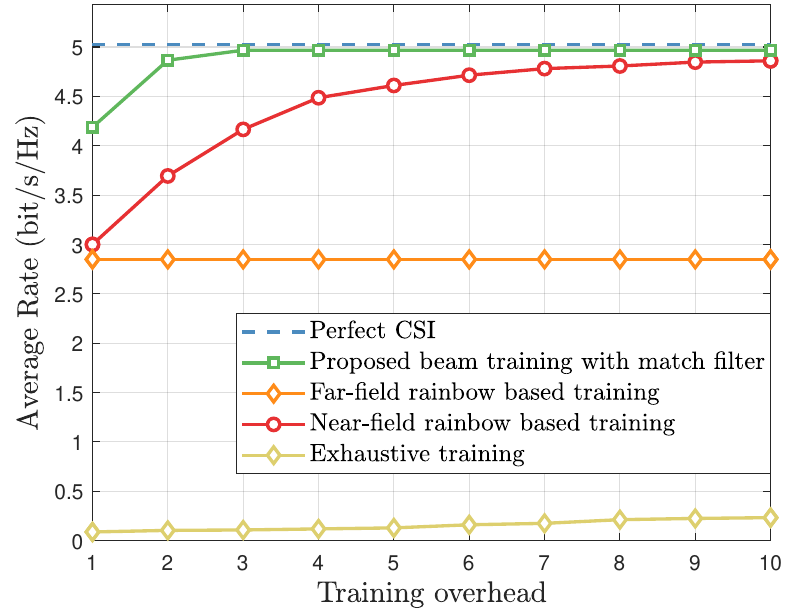}} 
	\subfigure[]{
		\includegraphics[width=0.95\linewidth]{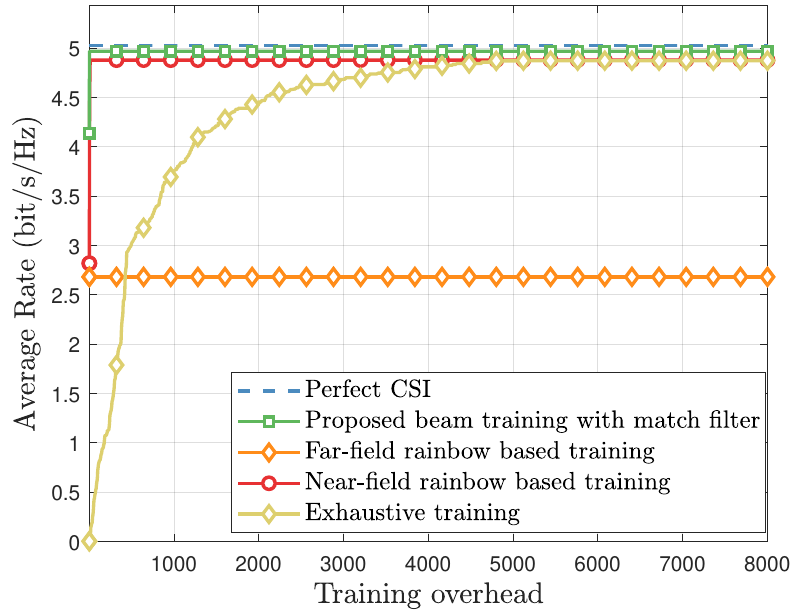}}
	\caption{Average rate performance vs. training overhead.}
	\label{pic_over}
\end{figure}
\else 
\begin{figure}
	\centering 
	\subfigure[]{
		\includegraphics[width=0.45\linewidth]{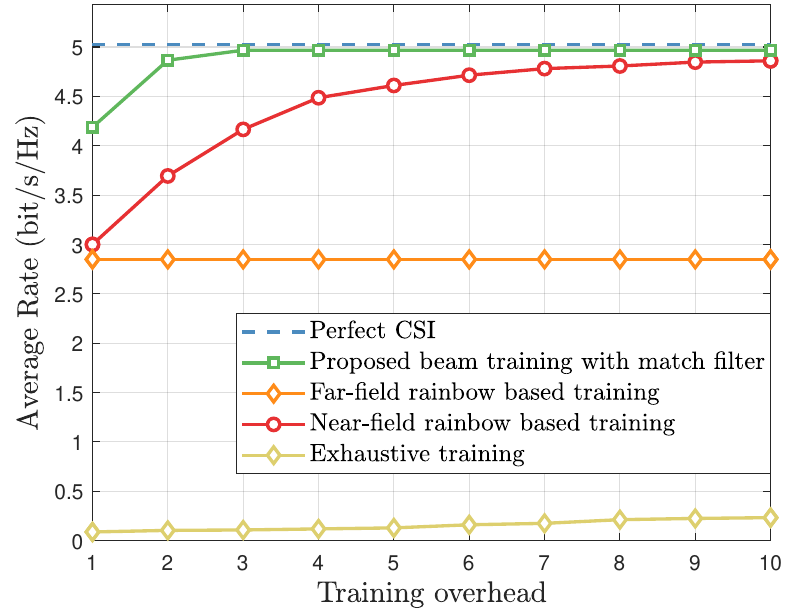}} 
	\subfigure[]{
		\includegraphics[width=0.45\linewidth]{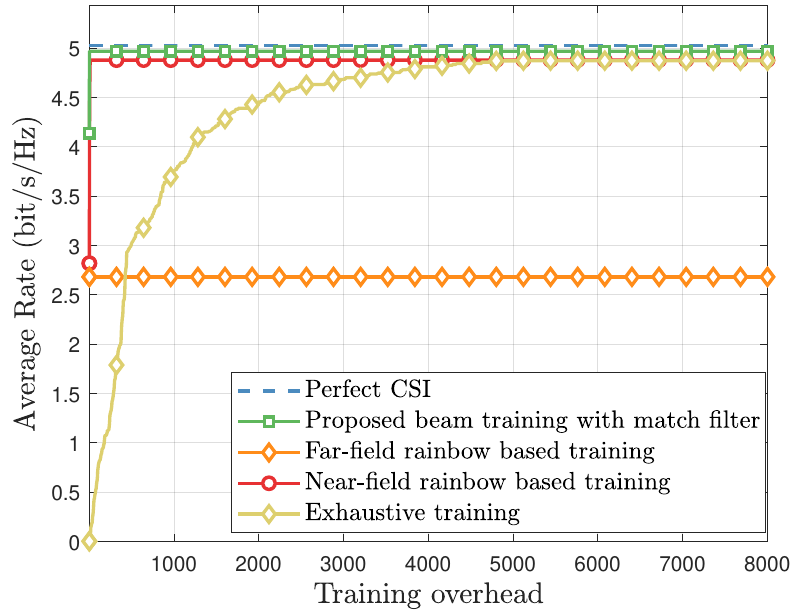}}
	\caption{Average rate performance vs. training overhead.}
	\label{pic_over}
\end{figure}
\fi

Fig.~\ref{pic_over} presents the average rate performance against the training overhead. The SNR is set as $15~{\rm dB}$. The training overhead ranges from $0$ to $8000$. In each time slot, we employ the optimal beamforming vector identified in prior time slots to serve the user. The far-field rainbow-based technique requires only one pilot. However, since the far-field approach focuses solely on angle information and disregards distance information, its average rate performance is suboptimal.
Furthermore,  near-field exhaustive scheme requires more than $2000$ pilots 
to attain a satisfactory performance, which is impractical for real-world systems.
Then we focus on the comparison of near-field rainbow-based beam training~\cite{nearrainbow} and the proposed method. For fair comparison, the number of sampled directions for the match filter is set as $L=1024$ to guarantee the same sample rate in the angle domain with near field rainbow ~\cite{nearrainbow} (the number of sampled directions is $M=1024$) while the number of sampled distances is both $S=10$. As depicted in Fig.~\ref{pic_over}, the proposed method employs only $3$ pilots to achieve comparable average rate performance while near-field rainbow-based method~\cite{nearrainbow} requires $10$. Therefore, since both the angle and distance domain can be searched with different subcarriers simultaneously, the proposed method further improves the near-field beam training efficiency.

\ifx\onecol\undefined
\begin{figure}
	\centering 
	\includegraphics[width=0.95\linewidth]{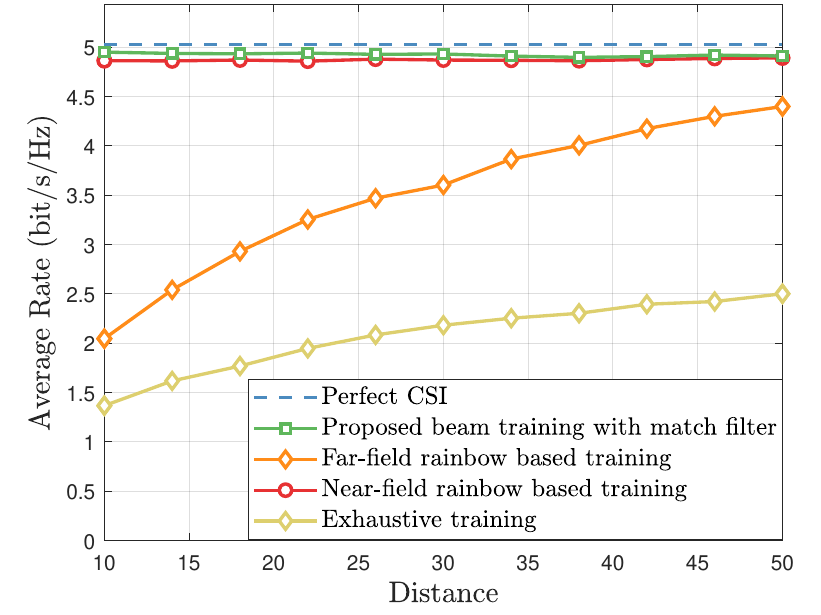}
	\caption{Average rate performance vs. distance.}
	\label{fig:snr_all}
\end{figure}
\else 
\begin{figure}
	\centering 
	\includegraphics[width=0.5\linewidth]{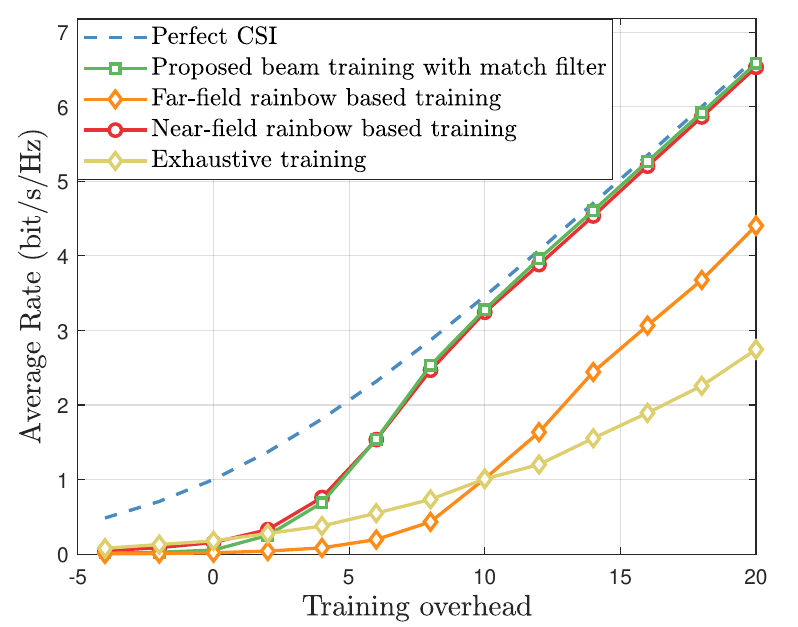}
	\caption{Average rate performance vs. distance.}
	\label{fig:snr_all}
\end{figure}
\fi

Fig.~\ref{fig:snr_all} illustrates the average rate performance versus the distance.
We have capped the maximum training overhead for all the methods under consideration as $256$ and set SNR as $15~{\rm dB}$. 
Additionally, the distance between the user and BS increases from $10$ to $50$ meters.
As illustrated in Fig.~\ref{fig:snr_all}, the proposed method with match filter achieves near-optimal rate performance with lower training overheads compared to near-field rainbow-based method.
Regarding the far-field scheme, as the distance decreases, near-field propagation becomes more prominent, leading to a rapid degradation in the average rate for the far-field scheme. 
Moreover, due to limited maximum training overhead, the exhaustive scheme
faces difficulties in searching for locations within the near field.


\section{Conclusions} \label{sec-con}
In this paper, we demonstrate a phenomenon called distance-dependent beam split where we can divide the subcarriers into different
groups which periodically cover the angular range with
different distance ranges by elaborately manipulation of the TD-PS parameters. Then a wideband beam training method has been proposed based on the new discovery and we further improve the performance for off-grid users. 
The simulation results have confirmed the efficacy of the proposed method, which emerges as a promising approach for achieving efficient beam training in XL-MIMO systems. The paper also presents the first attempt to fully unleash the capability of near-field beam split.
Future research could concentrate on extending the distance-dependent beam-split to uniform planar array configurations.
In addition,
investigation of the low-hardware-complexity TD-PS structure with a small number of TDs~\cite{twoleveltdd} is also left for future research.

\section*{Appendix} \label{appendix}
\subsection{Proof of \textbf{Lemma \ref{lem1_angle}}}\label{app:angle}
 Consider a beam focused on the location $(\theta,\alpha)$ at frequency $f$. To evaluate the beam coverage in the angle domain, we can assume that $\alpha=\bar{\alpha}$. Then, 
 the achieved beamforming gain on the location $(\bar{\theta},\bar{\alpha})$ is expressed  as 
	\begin{equation} \label{gain_theta}
		G(k(\theta-\bar{\theta}),0) =  \left\lvert \frac{\sin \frac{N_t \pi (\theta-\bar{\theta}) f }{2 f_c}}{N_t \sin \frac{\pi  (\theta-\bar{\theta}) f}{2 f_c}}  \right\rvert \overset{(a)}{\approx} \left\lvert \frac{\sin \frac{N_t \pi (\theta-\bar{\theta}) f }{2 f_c}}{ \frac{N_t \pi (\theta-\bar{\theta}) f }{2 f_c}}  \right\rvert, 
	\end{equation}
  where (a) holds because $N_t$ is very large. 
    In this case, $3 ~ {\rm dB}$-beam width in the angle domain can be numerically solved from $| G(k(\theta-\bar{\theta}),0)| = \frac{1}{\sqrt{2}}$:
    \begin{equation}
        |\theta-\bar{\theta}| \approx \frac{0.88f_c}{N_t f}.
    \end{equation} 
	To guarantee the entire coverage in the angle domain according to the reliability criterion, the angle difference of  $\theta_m$ and $\theta_{m+1}$ should be smaller than the sum of the $3 ~ {\rm dB}$-beam width of the beam at the $m$-th and $m+1$-th subcarrier, i.e., 
	\begin{equation}\label{eq_sumreq}
		|\theta_{m+1}-\theta_m|  \leqslant \eta_\theta^{f_m}+ \eta_\theta^{f_{m+1}} = \frac{0.88}{N_t}(\frac{f_c}{f_{m+1}}+\frac{f_c}{f_{m}}),
	\end{equation}
	where $\eta_\theta^{f}$ is the angle-domain $3 ~ {\rm dB}$-beam width of the beam at frequency $f$.
	Combining the above equation and (\ref{eq:dif}) we obtain
	\begin{equation}
		|f_c (\theta_p'+2p_m) (\frac{1}{f_{m+1}}-\frac{1}{f_m})| \leq \frac{0.88}{N_t}(\frac{f_c}{f_{m+1}}+\frac{f_c}{f_{m}}),
	\end{equation}
	which gives rise to the results of~\eqref{eq_t2} in {\bf Lemma \ref{lem1_angle}}.

\subsection{Derivation of the distance-domain 3 dB-beam width}

In~\eqref{eq_gain}, when $\theta=\bar{\theta}$, the array gain on an arbitrary physical distance $\bar{\alpha}$ is 
\ifx\onecol\undefined
\begin{align}\label{gain_alpha}
	G(0,k(\alpha-\bar{\alpha})) &= \frac{1}{N_t} \left\lvert \sum_{n=-N}^{N} e^{-jn^2d^2 k(\alpha-\bar{\alpha})} \right\rvert = F(\beta)
\end{align}
\else 
\begin{equation}\label{gain_alpha}
	G(0,k(\alpha-\bar{\alpha})) = \frac{1}{N_t} \left\lvert \sum_{n=-N}^{N} e^{-jn^2d^2 k(\alpha-\bar{\alpha})} \right\rvert 
= F(\beta)
\end{equation}
\fi
where $F(\beta) \overset{\Delta}{=} |C(\beta) + jS(\beta)|/\beta$, and  $C(\beta)=\int_0^\beta \cos(\frac{\pi}{2}t^2) {\rm d}t$ and $S(\beta)=\int_0^\beta \sin(\frac{\pi}{2}t^2) {\rm d}t$ are Fresnel functions.  The variable $\beta$ is written as
	$\beta = \sqrt{\frac{N_t^2d^2}{2\lambda}\left(\alpha-\bar{\alpha}\right)}$~\cite{nearCE}.
For $3 ~ {\rm dB}$-beam width, it can be numerically solved from $|F(\beta)| = \frac{1}{\sqrt{2}}$ that $\beta\approx 1.318$. Thus, the distance-domain $3 ~ {\rm dB}$-beam width at frequency $f$ can be denoted as
\begin{equation}
	\eta_\alpha^f = |\alpha-\bar{\alpha}| = \frac{\beta^2\lambda }{N_t^2 d^2} = \frac{4 \beta^2 f_c^2 }{N_t^2 c f}. 
\end{equation}

\subsection{Proof of \textbf{Proposition \ref{pro1_distance}}}\label{app:distance}
    We use the product the angle-domain array gain in \eqref{gain_theta} and distance-domain array gain in \eqref{gain_alpha} to approximate the array gain~\eqref{eq_gain} on an arbitrary physical location $(\bar{\theta},\bar{\alpha})$ as
        \begin{equation}\label{eq_g2}
            G(k(\theta-\bar{\theta}),k(\alpha-\bar{\alpha})) \approx \Xi_{N_t}\left(\frac{(\theta-\bar{\theta}) f}{ f_c}\right) F(\beta).
        \end{equation}
	Based on Taylor expansion, if $\bar{\theta},\bar{\alpha}$ is within $3 ~ {\rm dB}$-beam width of mainlobe (i.e. $\theta-\bar{\theta}\rightarrow 0, \alpha-\bar{\alpha}\rightarrow 0$), two parts of $G$ can be written as
	\begin{align}
		\Xi_{N_t}\left(\frac{(\theta-\bar{\theta}) f}{ f_c}\right) &\approx 1- \frac{1}{24}N_t^2 \pi^2 \frac{f^2}{f_c} (\theta-\bar{\theta})^2 \\
		F(\beta) &\approx  1- \frac{\pi^2}{90} \frac{N_t^4 d^4}{\lambda^2} (\alpha-\bar{\alpha})^2. \label{eq_Fapprox}
	\end{align}
	Combining~\eqref{eq_g2}-\eqref{eq_Fapprox} and neglecting the high-order small numbers, we arrive at~\eqref{eq_ell}.

\footnotesize

\bibliographystyle{IEEEtran}
\bibliography{rainbow, IEEEabrv}

\normalsize



\end{document}